\documentclass[11pt,twoside]{article}
\usepackage{amsmath}
\usepackage{amssymb}
\usepackage{amscd}
\usepackage{graphicx}
\newtheorem{definition}{Definition}[section]
\newtheorem{proposition}{Proposition}[section]
\newcommand{\remark}{\smallbreak\noindent{\bf Remark.}}
\renewcommand{\a}{\alpha}
\renewcommand{\b}{\beta}
\newcommand{\g}{\gamma}

\renewcommand{\d}{\delta}
\newcommand{\ze}{{\zeta}}
\renewcommand{\th}{\theta}
\newcommand{\Th}{\Theta}
\renewcommand{\l}{\lambda}
\newcommand{\m}{\mu}
\newcommand{\n}{\nu}
\renewcommand{\r}{\rho}
\newcommand{\s}{\sigma}
\renewcommand{\t}{\tau}
\newcommand{\om}{{\omega}}
\newcommand{\Cs}{{\rlap{\lower3pt\hbox{\textnormal{\LARGE \char'040}}}{\Gamma}}{}}
\newcommand{\de}{\partial}
\newcommand{\Eo}{{\scriptstyle{\mathrm{E}}}}
\newcommand{\detg}{{{\scriptstyle|}g{\scriptstyle|}}}
\newcommand{\rdg}{{\textstyle\sqrt{{\scriptstyle|}g{\scriptstyle|}}\,}}
\newcommand{\rrdg}{{\scriptstyle\sqrt{{\sst|}g{\sst|}}}}
\newcommand{\rdetg}{{\textstyle\sqrt{{\scriptstyle|}\det g{\scriptstyle|}}\,}}
\newcommand{\oh}{\tfrac{1}{2}}
\newcommand{\ih}{\tfrac{\iO}{2}}
\newcommand{\dde}[2]{\frac{\partial #1}{\partial #2}}
\newcommand{\cj}[1]{\overline{#1}}
\renewcommand{\.}{{\scriptstyle\boldsymbol{\dot{}}}}
\newcommand{\td}{\tilde}
\newcommand{\cev}[1]{\smash{\overset{\smash{{}_{\gets}}}{#1}}}
\newcommand{\myvec}[1]{\smash{\overset{\smash{{}_{\to}}}{#1}}}
\newcommand{\Lll}{{\scriptscriptstyle{\mathrm{L}}}}
\newcommand{\Rrr}{{\scriptscriptstyle{\mathrm{R}}}}
\newcommand{\Ttt}{{\scriptscriptstyle{\mathrm{T}}}}
\newcommand{\free}{{}_{\sst\mathrm{free}}}
\newcommand{\sbot}{{\scriptscriptstyle\bot}}
\newcommand{\sbo}{{\!\sbot}}
\newcommand{\spar}{{\scriptscriptstyle\|}}
\newcommand{\bz}{{\bar z}}
\newcommand{\bze}{{\bar\zeta}}
\newcommand{\bps}{{\bar\psi}}
\newcommand{\E}{{\boldsymbol{E}}}
\newcommand{\F}{{\boldsymbol{F}}}
\newcommand{\GA}{{\boldsymbol{\Gamma}}}
\renewcommand{\H}{{\boldsymbol{H}}}
\newcommand{\M}{{\boldsymbol{M}}}
\newcommand{\N}{{\boldsymbol{N}}}
\renewcommand{\P}{{\boldsymbol{P}}}
\newcommand{\Pm}{\P_{\!\!m}}
\newcommand{\T}{{\boldsymbol{T}}}
\newcommand{\Tt}{{\scriptscriptstyle{\boldsymbol{T}}}}
\newcommand{\U}{{\boldsymbol{U}}}
\newcommand{\Uc}{\cj{\U}}
\newcommand{\V}{{\boldsymbol{V}}}
\newcommand{\Vc}{\cj{\V}}
\newcommand{\W}{{\boldsymbol{W}}}
\newcommand{\Wc}{\cj{\W}}
\newcommand{\X}{{\boldsymbol{X}}}

\newcommand{\Z}{{\boldsymbol{Z}}}
\newcommand{\Zc}{\cj{\Z}}
\newcommand{\Lie}{\mathfrak{L}}
\newcommand{\CC}{{\mathbb{C}}}
\newcommand{\LL}{{\mathbb{L}}}
\newcommand{\NN}{{\mathbb{N}}}

\newcommand{\RR}{{\mathbb{R}}}
\newcommand{\UU}{{\mathbb{U}}}
\newcommand{\ZZ}{{\mathbb{Z}}}
\newcommand{\RRr}{{\scriptscriptstyle{\mathbb{R}}}}
\newcommand{\Ccal}{{\mathcal{C}}}
\newcommand{\Dcal}{{\mathcal{D}}}
\newcommand{\Ecal}{{\mathcal{E}}}
\newcommand{\Fcal}{{\mathcal{F}}}
\newcommand{\Hcal}{{\mathcal{H}}}
\newcommand{\Jcal}{{\mathcal{J}}}
\newcommand{\Kcal}{{\mathcal{K}}}
\newcommand{\Lcal}{{\mathcal{L}}}
\newcommand{\Mcal}{{\mathcal{M}}}
\newcommand{\Ncal}{{\mathcal{N}}}
\newcommand{\Ocal}{{\mathcal{O}}}
\newcommand{\Pcal}{{\mathcal{P}}}
\newcommand{\Qcal}{{\mathcal{Q}}}
\newcommand{\Tcal}{{\mathcal{T}}}
\newcommand{\Vcal}{{\mathcal{V}}}
\newcommand{\Zcal}{{\mathcal{Z}}}
\newcommand{\lfr}{\mathfrak{l}}

\newcommand{\DC}{{\boldsymbol{\Dcal}}}
\newcommand{\EC}{{\boldsymbol{\Ecal}}}
\newcommand{\FC}{{\boldsymbol{\Fcal}}}
\newcommand{\HC}{{\boldsymbol{\Hcal}}}
\newcommand{\LC}{{\boldsymbol{\Lcal}}}
\newcommand{\OC}{{\boldsymbol{\Ocal}}}
\newcommand{\QC}{{\boldsymbol{\Qcal}}}
\newcommand{\VC}{{\boldsymbol{\Vcal}}}
\newcommand{\ZC}{{\boldsymbol{\Zcal}}}
\newcommand{\DCo}{\DC_{\!\circ}}

\newcommand{\DCh}{{\rlap{\;/}\DC}}
\newcommand{\uDCh}{{\rlap{\;/}{\underline\DC}}}
\newcommand{\DCho}{{\rlap{\;/}\DCo}}
\newcommand{\VCo}{{\VC\!_{\circ}}}
\newcommand{\ZCo}{{\ZC_{\circ}}}
\newcommand{\uZC}{\underline\ZC}
\newcommand{\End}{\operatorname{End}}
\newcommand{\Tr}{\operatorname{Tr}}
\newcommand{\ad}{\operatorname{ad}}
\newcommand{\id}{{1\!\!1}}
\newcommand{\dO}{\mathrm{d}}
\newcommand{\dH}{\mathrm{d}_{\sst{\mathrm H}}}
\newcommand{\nasl}{{\rlap{\raise1pt\hbox{\,/}}\nabla}}
\newcommand{\GO}{\mathrm{G}}
\newcommand{\HO}{\mathrm{H}}
\newcommand{\LO}{\mathrm{L}}
\newcommand{\jO}{\mathrm{j}}
\newcommand{\JO}{\mathrm{J}}
\newcommand{\TO}{\mathrm{T}}
\newcommand{\TS}{\TO^{*}\!}
\newcommand{\VO}{\mathrm{V}}
\newcommand{\eO}{\mathrm{e}}
\newcommand{\iO}{\mathrm{i}}
\newcommand{\na}{\nabla\!}
\newcommand{\ten}[1]{\operatorname*{\otimes}_{\!{\scriptscriptstyle #1}} }
\newcommand{\cart}[1]{\operatorname*{\times}_{\!{\scriptscriptstyle #1}} }
\newcommand{\dir}[1]{\operatorname*{\oplus}_{\!{\scriptscriptstyle #1}} }
\newcommand{\we}{{\,\wedge\,}}
\newcommand{\weu}[1]{{\wedge^{\!#1}}}
\newcommand{\ve}{{\,\vee\,}}
\newcommand{\mdots}{{\cdot}{\cdot}{\cdot}}
\newcommand{\pint}{\mathord{\rfloor}}
\newcommand{\comp}{\mathbin{\raisebox{1pt}{$\scriptstyle\circ$}}}
\newcommand{\tn}{{\,\otimes\,}}
\newcommand{\bang}[1]{{\langle#1\rangle}}
\newcommand{\ket}[1]{{|#1\rangle}}
\newcommand{\Ii}[2]{{}^{#1}_{\phantom{#1}\!#2}}
\newcommand{\iI}[2]{{}_{#1}^{\phantom{#1}\!#2}}
\newcommand{\iIi}[3]{{}_{#1\phantom{#2}\!\!#3}^{\phantom{#1}\!#2}}
\newcommand{\sA}{{\scriptscriptstyle A}}
\newcommand{\sB}{{\scriptscriptstyle B}}
\newcommand{\sH}{{\scriptscriptstyle H}}
\newcommand{\sI}{{\scriptscriptstyle I}}
\newcommand{\sJ}{{\scriptscriptstyle J}}

\newcommand{\Bsf}{{\mathsf{B}}}
\newcommand{\bb}{{\mathsf{b}}}
\newcommand{\Xsf}{{\mathsf{X}}}
\newcommand{\ee}{{\mathsf{e}}}
\newcommand{\p}{{\scriptstyle\Pi}}
\newcommand{\uu}{{\mathsf{u}}}
\newcommand{\vv}{{\mathsf{v}}}
\newcommand{\sref}[1]{\S\ref{#1}}
\newcommand{\ie}{i.e$.$}
\newcommand{\eg}{e.g$.$}
\newcommand{\sst}{\scriptscriptstyle}
\newcommand{\qRq}{{\quad\Rightarrow\quad}}
\newcommand{\onto}{\rightarrowtail}
\newcommand{\FL}{\F_{\!\!\Lll}} \newcommand{\FR}{\F_{\!\!\Rrr}}
\newcommand{\gh}{\omega}
\newcommand{\ghost}{{}_{\sst\mathrm{ghost}}}
\newcommand{\spec}[1]{{}_{\sst{\mathrm{#1}}}}
\newcommand{\abs}[1]{{\mathrm{#1}}}
\newcommand{\cre}[1]{\abs{#1}^{\dag}}
\newcommand{\tra}[1]{\abs{#1}^{\!*}}
\newcommand{\brstQ}{{\scriptstyle{\mathrm Q}}}
\newcommand{\brstS}{{\scriptstyle{\mathrm S}}}
\newcommand{\suc}[1]{{\{\![#1]\!\}}}
\newcommand{\Suc}[1]{{\bigl\{\!\!\bigl[#1\bigr]\!\!\bigl\}}}
\newcommand{\swe}{\,{\scriptstyle\lozenge}\,}
\newcommand{\grade}[1]{{\lfloor#1\rceil}}
\newcommand{\scc}{{\mathsf{c}}}
\newcommand{\sco}[3]{\scc\Ii{#1}{#2#3}}
\newcommand{\gradezero}{{\!\sst\grade{\!0\!}}}
\newcommand{\gradeone}{{\!\sst\grade{\!1\!}}}
\newcommand{\lde}{\myvec{\de}}
\newcommand{\rde}{\cev{\de}}
\newcommand{\ldea}{{\rlap{\lower3pt\hbox{${\sst\sim}$}}\myvec{\de}}{}}
\newcommand{\rdea}{{\rlap{\lower3pt\hbox{${\sst\sim}$}}\cev{\de}}{}}
\title{On the geometry of ghosts}
\date{{\small June 8, 2015} }
\author{Daniel Canarutto\\[6pt]
{\small\it Dipartimento di Matematica e Informatica ``U.~Dini'', }\\
{\small\it Via S. Marta 3, 50139 Firenze, Italia}\\
{\small email:~daniel.canarutto@unifi.it}\\
{\small http://www.dma.unifi.it/\char126 canarutto}}
\begin{document}
\bibliographystyle{alpha}
\maketitle \thispagestyle{empty}
\begin{abstract}\noindent
An inspection of the precise geometric constructions
underlying fundamental notions in quantum gauge field theories
sheds light on various aspects which tend to be obscured
in the usual formalisms.
Revising the notions of mutually conjugated ``internal'' bundles
we propose a general rule for the constructions of free quantum fields
and their conjugates,
naturally yielding the needed fundamental properties
with regard to contractions, super-commutators, field momentum and Hamiltonian,
and other quantities.
This scheme applies to fields of all types;
in particular we examine consequences in relation to ghosts and anti-ghosts.
Finally in the context of observer-independent Fr\"olicher-smooth quantum bundles
we show how the antifield sectors naturally arise,
and examine the precise relation among these and the BRST symmetry
of a gauge field theory.
\end{abstract}

\bigbreak

\bigbreak

\vfill\newpage
\tableofcontents
\thispagestyle{empty}

\vfill\newpage

\setcounter{page}{1}

\section*{Introduction}

Notions in quantum field theory are often introduced in a matrix language,
which may obscure their geometric meaning
and blur differences among distinct objects,
while mathematically-oriented presentations tend to focus
on functional analytical aspects in Hilbert spaces~\cite{DeGe}.
Here we consider a somewhat different point of view,
believing that careful considerations in an explicitly geometric language
can help to clarify the matter further.
We can divide this task into various steps.
The first step consists of describing the underlying
finite dimensional bundles and their fiber structures,
namely the theory's ``classical'' (or ``pre-quantum'') setting.
While this aspect is widely treated in the literature,
for a deeper insight there are fine points deserving special consideration,
in particular with regard to the relation between mutually conjugate spaces,
the geometry of spinors and its connection with spacetime geometry,
and the geometry of pre-quantum ghost/anti-ghost fields.
Details about these points, according to the view of this paper,
can be found in previous work on a partly original presentation of
gauge field theories~\cite{C98,C00b,C07,C10a,C14a,C14c}.

As for the functional spaces suitable for describing quantum states,
we use an approach based on the distributional spaces of generalized semi-densities,
whence one constructs multi-particle state spaces
and a well-defined operator algebra generated by absorption and emission operators.
The notion of a free field can then be introduced
in a specialized context requiring a choosen observer in flat spacetime.
This approach can be extended to curved spacetime by regarding it
as a possibly local linearization associated with
a timelike submanifold (representing a \emph{detector}).
So we can see quantum fields, at least locally,
as sections of classical bundles tensorialized by a suitable
$\ZZ_2$-graded algebra $\OC$.
Though this construction depends on the chosen detector,
we may switch to a complementary view in which $\OC$-valued
fields over spacetime are considered as the fundamental objects;
so we ``forget'' about detectors and observers
and obtain a fully covariant,
observer-independent field theory\;---\;%
with the drawback that the notions of quantum states
and transition probabilities become blurred.
The differential geometric setting for the \emph{quantum bundles}
arising in this approach can be formulated in terms
of Fr\"olicher's notion of smoothness~\cite{C14b,Fr,FK,KM,CJK06,KolarModugno98}.

Free quantum fields are of special interest because,
together with point interactions,
they constitute (roughly speaking) the ``building blocks'' of field dynamics.
A close inspection of the construction
allows us to note details which are usually skipped.
A free quantum field is  best seen as a combination
of particle absorption and anti-particle emission;
this is true in all cases\,---\,not only for the Dirac field\,---\,%
except when the bundle of ``internal degrees of freedom'' is real
(then a particle and its anti-particle coincide).
Moreover, mutually conjugate fields are best seen
as analogous constructions on mutually conjugate bundles,
rather than equivalent objects obtained from one another by complex conjugation
(possibly associated with Hermitian transposition).
Indeed it is often remarked that the ghost and anti-ghost fields
are mutually independent.
We argue that this is actually true in general,
even when the Lagrangian and the field equations are preserved by conjugation.
Furthermore we exhibit a general rule for constructing the conjugate free fields
both in the boson and fermion cases,
and argue that the standard Dirac field $\bar\psi$ obeys it
though that is not explicitely shown in the literature.

Then we consider a generic gauge field theory whose sectors include
matter fermions, a gauge field, ghosts and anti-ghosts.
We write down their respective free-field expressions
according to the above said rule,
and check that several expected super-commutation identities
are indeed obeyed in all cases.
Moreover we compute the momentum and the Hamiltonian
for the free Dirac field and for the free ghost/anti-ghost field.
In the former case we recover the standard results,
in the latter case we find similar, reasonable results,
from which we argue that our construction is indeed sound.
We also compute the Faddeev-Popov current along free fields
in terms of emission and absorption operators.

In the last section we elaborate on the construction
of the quantum bundles suitable for an observer-independent field theory,
expanding on previous work~\cite{C14b}.
We construct the algebra of fiber polynomials,
and show how the antifield sectors and a Batalin-Vilkovisky algebra
naturally arise in this context.
Finally we examine the precise relation between the antifield sectors
and the BRST symmetry of a Lagrangian gauge field theory.

\section{From classical geometry to free quantum fields}
\label{s:From classical geometry to free quantum fields}
\subsection{Generalized semi-densities and quantum states}
\label{ss:Generalized semi-densities and quantum states}
Let \hbox{$\Z\onto\X$} be a finite-dimensional complex vector bundle
over the real $m$-dimensional orientable manifold $\X$,
and choose a ``positive'' semi-vector bundle
\hbox{$(\weu{m}\TS\X)^+\subset\weu{m}\TS\X$}.
Up to an isomorphism there is\footnote{
For an account of positive semi-spaces and their rational powers,
see~\cite{JMV10,C12a} and the bibliography therein.} 
a unique semi-vector bundle \hbox{$\UU\onto\X$}
such that \hbox{$\UU\tn\UU\cong(\weu{m}\TS\X)^+$}.
A section \hbox{$\X\to\UU\tn\Z$} is called
a \emph{$\Z$-valued semi-density}.
We denote as $\DCho(\X,\Z)$ the vector space of all such sections
which are smooth and have compact support.
Its dual space in the standard test map topology~\cite{Sc}
is indicated as \hbox{$\DCh(\X,\Z)$} and called the space of
\emph{$\Z$-valued generalised semi-densities}
(so the word ``generalized'' is used here in the distributional sense).
In particular, a sufficiently regular ordinary section
\hbox{$\th:\X\to\UU\tn\Z$} is in $\DCh(\X,\Z)$ via the rule
\hbox{$\bang{\th,\s}:=\int_\X\!\bang{\th(x),\s(x)}$}\,,
\hbox{$\s\in\DCho(\X,\Z^*)$}\,.

Semi-densities have a special status among all kinds of generalised sections
because of the natural inclusion $\DCho(\X,\Z)\subset\DCh(\X,\Z)$\,.
Furthermore, if a fibered Hermitian structure
of \hbox{$\Z\onto\X$} is assigned then one has the space $\LC^2(\X,\Z)$
of all ordinary semi-densities $\th$ such that
\hbox{$\bang{\th^\dag,\th}<\infty$}\,.
Let \hbox{$\boldsymbol{0}\subset\LC^2(\X,\Z)$} denote the subspace
of all almost-everywhere vanishing sections;
then the quotient \hbox{$\HC(\X,\Z)=\LC^2(\X,\Z)/\boldsymbol{0}$} is a Hilbert space,
and we get a so-called \emph{rigged Hilbert space}~\cite{BLT}
$$\DCho(\X,\Z)\subset\HC(\X,\Z)\subset\DCh(\X,\Z)~.$$
Elements in \hbox{$\DCh(\X,\Z)\setminus\HC(\X,\Z)$} can then be identified
with the (\emph{non-normalizable}) \emph{generalised states}
of the common physics terminology.

Let $\d[x]$ be the \emph{Dirac density} on $\X$
with support $\{x\}$\,, \hbox{$x\in\X$}\,.
A generalised semi-density is said to be \emph{of Dirac type}
if it is of the form \hbox{$\d[x]\tn u\in\DCh(\X,\Z)$}
with\footnote{
In the generalized sense $\d[x]$ is valued into $\weu{m}\TS\X$,
so that $\d[x]\tn u$ is valued into \hbox{$\UU\tn\UU\tn\UU^*\cong\UU$}.
} 
\hbox{$u:\X\to\UU^*\tn\Z$}.
We define $\uDCh(\X,\Z)$ to be the space of all
\emph{finite linear combinations} of Dirac-type semi-densities.
An important result in the theory of distributions~\cite{Sc}
then implies that $\uDCh(\X,\Z)$ is dense in $\DCh(\X,\Z)$,
namely any generalised semi-density can be approximated
with arbitrary precision
(in the sense of the topology of distributional spaces)
by a finite linear combination of Dirac-type semi-densities.

The assignment of a volume form \hbox{$\eta:\X\to(\weu{m}\TS\X)^+$}
and of a frame\footnote{
For notational simplicity we assume the frame's domain to be the whole $\X$.
} 
$\bigl(\bb_\a\bigr)$ of \hbox{$\Z\onto\X$}
determines the set \hbox{$\bigl(\Bsf_{x\a}\bigr)\subset\DCh(\X,\Z)$},
called a \emph{generalised basis},
where
$$\Bsf_{x\a}\equiv\d[x]\tn\eta^{-1/2}\tn\bb_\a(x)~.$$
Traditionally one would rather write $\Bsf_{x\a}$ as $\ket{x,\a}$ (say),
but the point here is that we can introduce a handy ``generalised index'' notation.
We write \hbox{$\Bsf^{x\a}\equiv\d[x]\tn\eta^{-1/2}\tn\bb^\a(x)$}\,,
where $\bigl(\bb^\a\bigr)$ is the dual classical frame.
Though contraction of any two distributions is not defined in the ordinary sense,
a straightforward extension of the discrete-space operation yields
$$\bang{\Bsf^{x'\!\a'},\Bsf_{x\a}}=\d^{x'}_x\,\d^{\a'}_{\a}~,$$
where $\d^{x'}_x$ is the generalised function
usually indicated as $\d(x'\,{-}\,x)$\,.
This is consistent with ``index summation'' in a generalised sense:
if \hbox{$z\in\DCho(\X,\Z)$}
and \hbox{$\ze\in\DCho(\X,\Z^*)$} are test semi-densities,
then we write
\begin{align*}
&z^{x\a}\equiv z^\a(x)\equiv\bang{\Bsf^{x\a},z}~,\quad
\ze_{x\a}\equiv \ze_\a(x)\equiv\bang{\ze,\Bsf_{x\a}}~,
\\[6pt]
&\bang{\ze,z}\equiv
\ze_{x'\!\a'}\,z^{x\a}\,\bang{\Bsf^{x'\!\a'},\Bsf_{x\a}}\equiv
\int_\X \ze_\a(x)\,z^\a(x)\,\eta(x)~,
\end{align*}
namely we interpret index summation
with respect to the continuous variable $x$ as integration,
provided by the chosen volume form.
This formalism can be extended to the contraction
of two generalised semi-densities whenever it makes sense.

\subsection{Multi-particle states and elementary operators}
\label{ss:Multi-particle states and elementary operators}

In order to deal with multi-particle states and different particle types,
we introduce further notations
in the context of~\sref{ss:Generalized semi-densities and quantum states}.
We set
\begin{align*}
&\ZCo\equiv\DCho(\X,\Z)~,\quad
\uZC^1\equiv\uDCh(\X,\Z)~,\quad
\ZC^1\equiv\DCh(\X,\Z)~,
\\[6pt]
&\uZC^n\equiv\lozenge^n\uZC^1~,\quad
\ZC^n\equiv\lozenge^n\ZC^1~,
\end{align*}
where $\lozenge$ denotes either symmetrized or antisymmetrised tensor product
(respectvely for bosons and fermions).
Then $\uZC^n$ turns out to be dense in $\ZC^n$,
which in turn is dense either in the symmetrised
or in the antisymmetrised subspace of $\DCh(\X^n,{\otimes}^n\Z)$\,,
\hbox{$\X^n\equiv\X\,{\times}\,\mdots\,{\times}\,\X$}\,.
Next we set \hbox{$\ZC\equiv\bigoplus_{n=0}^\infty\ZC^{n}$},
and assemble several particle types into one total state space
$$\VC:=\ZC'\tn\ZC''\tn\mdots\equiv\textstyle{\bigoplus_{n=0}^\infty}\VC^{n}~,$$
where $\VC^{n}$, constituted of all elements of tensor rank $n$\,,
is the space of all states of $n$ particles of any type.

We can also consider a dual construction, in an elementary sense,
by replacing $\Z$ with its dual $\Z^*$,
and obtain the ``dual'' space $\VC^{*}$.
Moreover we note that using test semi-densities
we obtain subspaces \hbox{$\VCo\subset\VC$}
and \hbox{$\VC\!_{\circ}^{\:{*}}\subset\VC^{*}$}.

If we now let the \emph{parity} (or \emph{grade}) $\grade{\phi}$
of a monomial element (a ``decomposable tensor'') \hbox{$\phi\in\VC$}
to be the number of its fermion factors $({\mathrm{mod}}\;2)$,
then we obtain on $\VC$ a structure of ``super-algebra'' (a $\ZZ_2$-graded algebra).
The algebra product, which we denote as $\scriptstyle\lozenge$,
is the tensor product modulo the so-called \emph{Koszul convention},
which essentially amounts to imposing anti-commutativity.
Furthermore we can consider ``interior products'',
in the appropriate tensor factors,
between elements in $\VC^{*}$ and elements in $\VC$,
possibly to be intended in a generalized sense.
These will be indicated by a vertical bar as (say) $\ze\,|\,\psi$\,.
We obtain the rules
\begin{align*}
&\psi\swe\phi=(-1)^{\grade{\phi}\grade{\psi}}\phi\swe\psi~,\quad
(\ze\swe\xi)\,|\,\psi=\xi\,|\,(\ze\,|\,\psi)~,
\\[6pt]
&\ze\,|\,(\phi\swe\psi)=
(\ze\,|\,\phi)\swe\psi+(-1)^{\grade{z}\grade{\phi}}\,\phi\swe(\ze\,|\,\psi)~,\qquad
\phi,\psi\in\VC,~
\ze,\xi\in\VC^{{*}1},
\end{align*}
valid whenever each of the involved factors has a definite parity.
A linear map \hbox{$X:\VC{}\to\VC{}$} is called
a \emph{super-derivation} (or \emph{anti-derivation}) \emph{of grade~$\grade{X}$}
if \hbox{$\grade{X\psi}=\grade{X}+\grade{\psi}$} and the graded Leibnitz rule
$$X(\phi\swe\psi)=(X\phi)\swe\psi+
(-1)^{\grade{X}\grade{\phi}}\phi\swe X\psi$$
is fulfilled.

The \emph{absorption} operator associated with \hbox{$\ze\in\VC^{{*}1}$}
and the \emph{emission} operator associated with \hbox{$z\in\VC^1$}
are the linear maps \hbox{$\VCo\to\VC$} respectively defined as
$$\abs{a}[\ze]\phi\equiv \ze\,|\,\phi~,\quad
\tra{a}[z]\phi\equiv z\swe\phi~,\qquad\phi\in\VC~.$$
Similarly, we have operators
\hbox{$\abs{a}[z],\tra{a}[\ze]:\VC_{\!\circ}^{*}\to\VC^{*}$},
and one easily checks that $\abs{a}[\ze]$ and $\tra{a}[\ze]$
are mutually transposed maps.
Absorption and emission operators generate a vector space
which turns out to be a $\ZZ_2$-graded algebra
(the algebra product being the composition of endomorphisms)
by letting the grades of $\abs{a}[\ze]$ and $\tra{a}[z]$
be $\grade{\ze}$ and $\grade{z}$\,, respectively.
The \emph{super-bracket} of two operators \hbox{$X,Y$}
in this space is then defined by
$$\suc{X,Y}:=X\,Y-(-1)^{\sst\grade{X}\grade{Y}}Y\,X~.$$
In particular, for \hbox{$y,z\in\VC^1$}
and \hbox{$\ze,\xi\in\VC^{{*}1}$} we get
$$\suc{\abs{a}[\xi],\abs{a}[\ze]}=\suc{\tra{a}[y],\tra{a}[z]}=0~,\quad
\suc{\abs{a}[\ze],\tra{a}[z]}=\bang{\ze,z}\,\id~.$$

The vector space $\OC^1$ of all sums of the kind \hbox{$\abs{a}[\ze]+\tra{a}[z]$}
has the subspace $\underline\OC^1$ of all finite linear combinations of
absorption and emission operators associated with Dirac-type semi-densities.
In particular we write \hbox{$\abs{a}^{x\a}\equiv\abs{a}[\Bsf^{x\a}]$}\,,
\hbox{$\tra{a}_{x\a}\equiv\tra{a}[\Bsf_{x\a}]$}\,,
and obtain super-commutation rules
$$\Suc{\abs{a}^{x\a},\abs{a}^{x'\!\a'}}=
\Suc{\tra{a}_{x\a},\tra{a}_{x'\!\a'}}=0~,\quad
\Suc{\abs{a}^{x\a},\tra{a}_{x'\!\a'}}=\d^\a_{\a'}\,\d^x_{x'}~,$$
where the latter is to be understood in a generalised sense:
for \hbox{$\ze\in\VC\!_{\circ}^{\;{*}1}$}, \hbox{$z\in\VC\!_{\circ}^{\;1}$}\,,
we write
$$\Suc{\abs{a}[\ze],\tra{a}[z]}=
\Suc{\ze_{x\a}\,\abs{a}^{x\a},z^{x'\!\a'}\,\tra{a}_{x'\!\a'}}=
\ze_{x\a}\,z^{x'\!\a'}\,\Suc{\abs{a}^{x\a},\tra{a}_{x'\!\a'}}=
\bang{\ze,z}~.$$

Next we denote as $\OC^n$, \hbox{$n\in\NN$}\,,
the vector space spanned by all compositions of
$n$ emission and absorption operators ordered in such a way
that all absorption operators stand on the right of any emission operator
(\emph{normal order}).
A product \hbox{$\OC^n\times\OC^p\to\OC^{n+p}$} can be defined as composition
together with normal reordering,
obtained by imposing the \emph{modified rule}
$$\Suc{\abs{a}^{x\a},\tra{a}_{x'\!\a'}}=0~.$$
Setting \hbox{$\OC^0\equiv\CC$}
we obtain a graded algebra \hbox{$\OC\equiv\bigoplus_{n=0}^\infty\OC^{n}$}
of linear maps \hbox{$\VCo\to\VC$}
(note that \emph{normal ordering is needed} for obtaining an algebra of such maps).
Moreover, $\OC$ turns out to be a $\ZZ_2$-graded algebra,
which can be identified with $\VC\tn\VC^*$.

A suitable extension of $\OC$ will be actually needed.
Let \hbox{$Z:\RR\to\OC$} be a local curve such that
\hbox{$\lim_{\l\to0}[Z(\l)\chi]\in\VC$} exists in the sense of distributions
for all \hbox{$\chi\in\VCo$}\,.
Then $\lim_{\l\to0}Z(\l)$ is a well-defined linear map \hbox{$\VCo\to\VC$}
which belongs, in general, to an extended space \hbox{$\OC^\bullet\supset\OC$}\,.

\subsection{Conjugation and the role of Hermitian structure}
\label{ss:Conjugation and the role of Hermitian structure}

In order to understand the precise relation between mutually conjugate fields,
we need to keep in mind the notion of \emph{anti-dual space} $\Vc{}^*$
and of \emph{conjugate space} $\Vc$ of a finite-dimensional complex vector space $\V$.
In the finite-dimensional situation,
the former can be simply defined as the complex vector space of all anti-linear
functions \hbox{$\V\to\RR$}\,, and the latter as its dual space.
Complex conjugation yields then \emph{anti-isomorphims} \hbox{$\V\leftrightarrow\Vc$}
and \hbox{$\V^*\leftrightarrow\Vc{}^*$}.
The ``dotted-index'' formalism is useful for dealing with
component expressions related to $\Vc$ and $\Vc{}^*$.

The above notions can be seamlessly extended to complex vector bundles,
and we observe that in most practical cases
the fibers are assumed to be endowed with a Hermitian structure.
This yields various isomorphisms
and consequent possible simplifications of indexed expressions,
specially with regard to conjugation.
Nevertheless, a few preliminary hair-splitting observations
may help us to handle the ensuing formalism better.

A Hermitian structure on \hbox{$\Z\onto\X$} is a non-degenerate tensor field
$$h:\X\to\Zc{}^*\ten{\X}\Z^*$$
such that \hbox{$\bar h=h^\Ttt$}.
If $\bigl(\bb_\a\bigr)$ is a frame of $\Z$
then the conjugate frame of $\Zc$ is denoted as $\bigl(\bar\bb_{\a\.}\bigr)$\,,
and the anti-dual frame of $\Zc{}^*$
is denoted as $\bigl(\bar\bb^{\a\.}\,\bigr)$\,.
Accordingly we write
\begin{align*}
&h=h_{\a\.\,\a}\,\bar\bb^{\a\.}\tn\bb^\a~,&&
h^\#=h^{\a\.\,\a}\,\bar\bb_{\a\.}\tn\bb_\a~,
\\[6pt]
&h_{\a\.\,\a}\,h^{\a\.\,\b}=\d^\b_\a~,&&
h_{\a\.\,\a}\,h^{\b\,\.\,\a}=\d^{\b\,\.}_{\a\.}~,
\end{align*}
where \hbox{$h^\#:\X\to\Zc\ten{\X}\Z$} is the ``inverse'' of $h$\,.
Then $h$ and $h^\#$ determine isomorphisms
\begin{align*}
&\flat:\Zc\to\Z^*:\bz\mapsto\bz^\flat~,&&
\bar\flat:\Z\to\Zc{}^*:z\mapsto z^\flat~,
\\[6pt]
&\#:\Zc{}^*\to\Z:\bze\mapsto\bze^\#~,&&
\bar\#:\Z^*\to\Zc:\ze\mapsto\ze^\#~,
\end{align*}
over $\X$, with $\flat$ and $\bar\#$ being mutually inverse
as well as $\bar\flat$ and $\#$\,.

If \hbox{$z=z^\a\,\bb_\a\in\Z$}, \hbox{$\ze=\ze_\a\,\bb^\a\in\Z^*$},
then we also write \hbox{$\bz=\bz^{\a\.}\,\bar\bb_{\a\.}\in\Zc$},
\hbox{$\bze=\bze_{\a\.}\,\bar\bb^{\a\.}\in\Zc{}^*$}, and
\begin{align*}
&\bz^\flat=h_{\a\.\,\a}\,\bz^{\a\.}\,\bb^\a\equiv\bz_\a\,\bb^\a~,
&& z^\flat=h_{\a\.\,\a}\,z^\a\,\bar\bb^{\a\.}\equiv z^\a\,\bar\bb_\a~,\\[6pt]
&\ze^\#=h^{\a\.\,\a}\,\ze_\a\,\bar\bb_{\a\.}\equiv \ze_\a\,\bar\bb^\a~,
&& \bze^\#=h^{\a\.\,\a}\,\bze_{\a\.}\,\bb_\a\equiv\bze^\a\,\bb_\a~.
\end{align*}
Thus in many cases the Hermitian structure allows avoiding ``dotted indices'',
which are often used to distinguish components in conjugate spaces.
In particular we may use
$$\quad \bb_\a^\flat=\bar\bb_\a\equiv h_{\a\.\,\a}\,\bar\bb^{\a\.}~,\qquad
\bb^{\a\#}=\bar\bb^\a\equiv h^{\a\.\,\a}\,\bar\bb_{\a\.}~.$$

A Hermitian structure is specially relevant in relation to the fact
that whenever a sector corresponding to a complex bundle $\Z$ is considered,
then the theory also includes the sector corresponding to the conjugate bundle $\Zc$.
These two classical bundles underlie the description
of a couple particle-antiparticle.\footnote{
Simplifications ensue if \hbox{$\Z\equiv\CC\tn\Z_{\!\RRr}$}
where $\Z_{\!\RRr}$ is a real vector bundle.
Also note that a real metric $g$ of $\Z_{\!\RRr}$
can be naturally extended to the a Hermitian structure
of the complexified bundle.} 
Accordingly, from an ordinary section \hbox{$\ze:\X\to\Z^*$}
we get operators $\abs{a}[\ze]$ \emph{and} $\tra{a}[\ze^\#]$\,,
which can be respectively seen as the absorption of a particle
and the emission of the related anti-particle.
Similarly, an ordinary section \hbox{$z:\X\to\Z$} yields
operators $\tra{a}[z]$ and $\abs{a}[z^\flat]$\,.
Extending this construction to generalized sections,
and in particular to the elements
of the generalized frame $\bigl(\Bsf_{x\a}\bigr)$
and of the generalized dual frame $\bigl(\Bsf^{x\a}\bigr)$,
we get the elementary anti-particle emission and absorption operators
$$\tra{a}{}^{x\a}\equiv\tra{a}[(\Bsf^{x\a})^\#]~,\qquad
\abs{a}_{x\a}\equiv\abs{a}[(\Bsf_{x\a})^\flat]~.$$

We stress that, like the correspondences \hbox{$\ze\to\ze^\#$}
and \hbox{$z\to z^\flat$} \emph{do not imply conjugation of $\ze$ and $z$},
so do $\tra{a}{}^{x\a}$ and $\abs{a}_{x\a}$\,.
Actually, the conjugation relation is between the spaces on which these operators act,
not between the operators themselves as generalized functions of $x$\,.
Hence we do not use the common notation $\cre{a}$ for emission operators,
as the implied usual meaning for the ``dagger'' label
is ``transposition together with conjugation''.
On the other hand, we could rightly set
\hbox{$\cre{a}[\ze]\equiv\tra{a}[\bar\ze{}^\#]$}\,,
so that we see that if the Hermitian structure is positive-definite
\emph{and} only orthonormal frames are considered,
then that notation doesn't create difficulties
essentially because one may identify high and low indices
as well as dotted and non-dotted indices.
Such identifications are routinely made in the literature~\cite{DeGe,IZ}.

As for the super-commutation rules for the two new elementary operators
(besides the rules stated in~\sref{ss:Multi-particle states and elementary operators}),
we note that for ordinary sections $z$ and $\ze$ we have
$$\Suc{\abs{a}[z^\flat],\tra{a}[\ze^\#]}=
\bang{z^\flat,\ze^\#}\,\id=\bang{\ze,z}\,\id~,$$
so that we get the generalized identity
$$\suc{\abs{a}^\a(x)\,,\,\tra{a}_\b(y)}=
\suc{\abs{a}_\b(x)\,,\,\tra{a}{}^\a(y)}=
\d^\a_\b\,\d(x-y)~,$$
which is independent of the signature of the Hermitian structure $h$\,.

Other super-commutators vanish.
In particular, we note that an emission operator and an absorption
operator with the same index type are related to internal states
of a particle and its anti-particle,
which are in general distinct. Thus
$$\suc{\abs{a}_\a(x)\,,\,\tra{a}_\b(y)}=
\suc{\abs{a}^\a(x)\,,\,\tra{a}{}^\b(y)}=0~.$$

Finally, we note that allowing normal ordering amounts to assuming
the \emph{modified rules}
$$\suc{\abs{a}^\a(x)\,,\,\tra{a}_\b(y)}=
\suc{\abs{a}_\b(x)\,,\,\tra{a}{}^\a(y)}=0~.$$

\subsection{Distributional bundles and generalized frames}
\label{ss:Distributional bundles and generalized frames}

For a given particle type in Einstein's spacetime $(\M,g)$,
the underlying ``classical''
geometric structure is that of a 2-fibered bundle \hbox{$\Z\to\Pm\to\M$},
where the top fibers describe the ``internal degrees of freedom''
and \hbox{$\Pm\subset\P\cong\TS\M$}
is the sub-bundle over $\M$ of future shells for the particle's mass $m$\,.
At each \hbox{$x\in\M$} we perform the constructions presented
in the previous sections,
with the generic manifold $\X$ now replaced by $(\Pm)_x$\,.
In particular we get spaces \hbox{$\ZC^1_x\equiv\DCh((\Pm)_x,\Z\!_x)$}\,,
the fibered set \hbox{$\ZC^1:=\bigsqcup_{x\in\M}\!\!\ZC^1_x$}
and the \emph{multi-particle state} bundle
$$\ZC:=\textstyle{\bigoplus_{n=0}^\infty}\ZC^{n}\onto\M~.$$
It turns out that \hbox{$\ZC\onto\M$},
as well as other similar or related bundles,
is naturally a smooth vector bundle according to Fr\"olicher's notion
of smoothness~\cite{C14b,Fr,FK,KM,CJK06,KolarModugno98}.

Considering more particle types,
one eventually gets the total quantum bundle\footnote{
The quantum bundles for particle types of different mass
are constructed over different mass-shell bundles.} 
$$\VC:=\ZC'\tn\ZC''\tn\ZC'''\tn\mdots=
\textstyle{\bigoplus_{n=0}^\infty}\VC^{n}\onto\M~.$$
Similarly, one gets the Fr\"olicher-smooth vector bundles
\hbox{$\ZCo\onto\M$} of all test fiber semi-densities
and \hbox{$\uZC\onto\M$} of all finite sums fiber semi-densities of Dirac type.

Now consider an orthogonal splitting
\hbox{$\TS\M\equiv\P=\P_{\!\!\spar}\oplus\P_{\!\!\sbot}$}
into ``timelike'' and ``spacelike'' $g$-orthogonal subbundles over $\M$
(which can be seen as associated to the choice of an observer).
Let $\eta_\sbot$ be the volume form, associated with the metric,
on the fibers of \hbox{$\P_{\!\!\sbot}\onto\M$}.
The orthogonal projection
\hbox{$\P\to\P_{\!\!\sbot}$} yields a distinguished diffeomorphism
\hbox{$\Pm\leftrightarrow\P_{\!\!\sbot}$} for each $m$.
The pull-back of $\eta_\sbot$\,, denoted by the same symbol, is then
a volume form on the fibers of $\Pm$\,.
The \emph{Leray form}\footnote{
Let $\M$ be a manifold with a chosen volume form $\eta$\,,
and $f$ a function on $\M$ such that the submanifold \hbox{$\N\subset\M$}
is characterized by \hbox{$f=0$} and $\dO f$ nowhere vanishes on $\N$.
Then the Leray form $\om[f]$, often denoted as $\d(f)$\,,
is characterized~\cite{CdW} by the condition that \hbox{$\dO f\we\om[f]=\eta$}
holds on $\N$.} 
$$\om_m\equiv\om[p_0-\Eo_m(p_\sbo)]~,\qquad
\Eo_m(p_\sbo)=(m^2+|p_\sbo|^2)^{1/2}~,$$
can now be then written as
$$\om_m(p)=(2\,p_0)^{-1}\eta_\sbot(p)~,\quad
p\in\Pm~,\quad p_0\equiv\Eo_m(p_\sbo)~.$$
This is a distinguished $3$-form on each fiber of \hbox{$\Pm\onto\M$},
and can also be regarded as a generalized density on each fiber of $\P$.

It will be convenient to use the ``spatial part''
$p_\sbo$ of the 4-momentum $p$ as a label,
that is a generalised index for quantum states.
If $\bigl(\bb_\a\bigr)$ is a frame of \hbox{$\Z\onto\Pm$} then we consider
the generalised frame
\hbox{$\bigl\{\Bsf_{p\a}\bigr\}\equiv\bigl\{\Xsf_p\tn\bb_\a\bigr\}$}\,,
where $\Xsf_p$ is defined as follows.
For each \hbox{$p\in\Pm$} let $\d_m[p]$
the Dirac density with support $\{p\}$ on the same fiber of \hbox{$\Pm\onto\M$}\,,
and let \hbox{$\d(y_\sbo\!{-}p_\sbo)$} be the generalised function characterised by
\hbox{$\d_m[p](y_\sbo)=\d(y_\sbo\!{-}p_\sbo)\,\dO^3y_\sbo$}
in terms of linear coordinates
\hbox{$\bigl(y_\l\bigr)\equiv\bigl(y_0,y_1,y_2,y_3\bigr)
\equiv\bigl(y_0,y_\sbo\bigr)$}
in the fibers of $\P$.
Then for each \hbox{$p\in\Pm$} we regard $\Xsf_p$
as a generalised function of the variable $y_\sbo$\,,
with the expression
$$\Xsf_p(y_\sbo):=l^{-3/2}\,\d(y_\sbo{-}p_\sbo)\,\sqrt{\dO^3y_\sbo}~.$$
Here $l$ is a constant length needed
in order to get an unscaled (``conformally invariant'') semi-density.

\subsection{Quantum configuration space}
\label{ss:Quantum configuration space}

In order to build a viable theory of quantum particles and their interactions
one needs a time function, possibly associated with an observer of some kind.
Having a global such structure in curved spacetime is a non-trivial requirement.
However we may consider a somewhat weaker setting~\cite{C05,C12a},
based on the assignment of a \emph{detector},
that is a timelike submanifold \hbox{$\T\subset\M$};
indeed a momentum-space formalism for particle interactions,
in terms of generalised semi-densities,
can be exhibited as a sort of a complicated `clock' carried by it.
In the case of an inertial detector in flat spacetime,
the Fourier transform relates the momentum-space
and the position-space formalisms;
this correspondence can be naturally extended to the curved spacetime case but,
in general, only locally (in a sense to be made precise).

A \emph{generalised frame of free one-particle states}
along $\T$ can be introduced by fixing any event \hbox{$t_0\in\T\subset\M$}
and a classical frame
\hbox{$\bigl(\bb_\a\bigr)$} of the bundle \hbox{$\Z\onto(\Pm)_{t_0}$}\,.
The family of generalised semi-densities
$\bigl\{\Bsf_{p\a}(t_0)\bigr\}$
is then a generalised frame of \hbox{$\ZC^1_{t_0}\onto(\Pm)_{t_0}$}\,,
which can be transported along $\T$
by virtue of the underlying geometric structure.\footnote{
This includes Fermi transport~\cite{C09a,C12a}
for the spacetime related factors,
and a background connection of $\Z$
which will have to be assumed~\cite{C12a}.} 
We obtain sections
$$\Bsf_{p\a}:\T\to\DCh(\Pm,\Z)_\Tt:
t\mapsto\Bsf_{p\a}(t)=\Xsf_{p(t)}\tn\bb_\a~,$$
where \hbox{$p:\T\to\Pm:t\mapsto p(t)$} is Fermi-transported.
This yields a trivialization
$$\DCh(\Pm,\Z)_\Tt\cong\T\times\DCh(\Pm,\Z)_{t_0}~,$$
which can be seen as determined by a suitable connection called
the \emph{free-particle connection}.
Eventually, the above arguments can be naturally extended
to multi-particle bundles and states.
When several particle types are considered,
we get a trivialization \hbox{$\VC_{\!\Tt}\cong\T\times\QC$}
of the total quantum state bundle,
where \hbox{$\QC\equiv\VC_{\!t_0}$} can be seen as the
``quantum configuration space''.
The quantum interaction,
an added term that modifies the free-field connection,
can be constructed by assembling the classical interaction
with a distinguished quantum ingredient~\cite{C05,C12a}.
By construction, the free-particle transport
preserves particle type and number.
Accordingly, we also get the $\ZZ_2$-graded operator algebra
$$\OC\cong\QC\tn\QC^{*}\equiv\VC_{\!t_0}\tn\VC_{\!t_0}^{*}~,$$
where the identification is determined via normal ordering.

The relation to position-space formalism can be summarized as follows.
The restriction of the tangent bundle of $\M$ to base $\T$ splits as
\hbox{$(\TO\M)_{{}_\T}=
(\TO\M)_\Tt^{\spar}\oplus(\TO\M)_\Tt^{\sbot}$}
into ``timelike'' and ``spacelike'' $g$-orthogonal subbundles.
Exponentiation determines, for each $t\in\T$,
a diffeomorphism from a neighbourhood of $0$ in $(\TO\M)_t^{\sbot}$
to a spacelike submanifold $\M\!_t\subset\M$,
and so a 3-dimensional foliation of a neighbourhood
\hbox{$\N\equiv\bigcup_{t\in\T}\M\!_t\subset\M$} of $\T$.
A \emph{tempered} generalised semi-density on $(\Pm)_t$ yields,
via Fourier transform, a generalised semi-density on
\hbox{$(\TO\M)_t^{\sbot}$}\,.
A suitable restriction\footnote{
A distribution can be restricted to an open set~\cite{Sc}.} 
then yields, via exponentiation,
a generalised semi-density on $\M\!_t$\,.
This correspondence can be extended to $\Z$-valued semi-densities
by means of background linear connections of the various ``internal'' bundles.
Eventually, the trivialisation \hbox{$\VC_{\!\Tt}\cong\T\times\QC$}
can be extended as \hbox{$\VC_{\!\sst\N}\cong\N\times\QC$}\,.
For an inertial detector in flat spacetime
we essentially get the usual correspondence between
momentum-space and position-space representation.

\subsection{Free quantum fields}
\label{ss:Free quantum fields}

If a fibered Hermitian structure of \hbox{$\Z\onto\Pm$} is assumed,
then any \hbox{$\ze\in\DCh(\Pm,\Z^*)$} yields an absorption operator $\abs{a}[\ze]$
and an emission operator $\tra{a}[\ze^\#]$ as well.
Proceeding as in~\sref{ss:Conjugation and the role of Hermitian structure}
we can now see $\abs{a}^{\a}$ and $\tra{a}{}^{\a}$
as generalised functions of momentum,
which in terms of the previously described generalized frames
can be written as
$$\abs{a}^{\a}(p_\sbo)\equiv\abs{a}^{p\a}:=\abs{a}[\Bsf^{p\a}]\equiv
\abs{a}[\Xsf^p\tn\bb^\a]~,\quad
\tra{a}{}^{\a}(p_\sbo)\equiv\tra{a}{}^{p\a}:=
\tra{a}[(\Bsf^{p\a})^\#]\equiv\tra{a}[\Xsf^p\tn\bar\bb^\a]~.$$
Consistently with the generalised index notation we also write
\hbox{$\abs{a}[\ze]=\ze_{p\a}\,\abs{a}^{p\a}$},
\hbox{$\tra{a}[\ze]=\ze_{p\a}\,\tra{a}{}^{p\a}$},
and eventually
$$\abs{a}[\_]=\abs{a}^{p\a}\,\Bsf_{p\a}~,\quad
\tra{a}[\_]=\tra{a}{}^{p\a}\,\Bsf_{p\a}~.$$

Essentially, free quantum fields are introduced as combinations
of Fourier transforms and anti-transforms of the above objects.
However, the fact that $\Z$ is in general a vector bundle over $\Pm$
may stand in the way of expressing a quantum field
as a section of some bundle over $\M$.
In order to overcome this difficulty we first note that
in the situations of interest $\Z$ is a subbundle
of a ``semi-trivial'' bundle, namely
$$\Z\subseteq\Pm\cart{\M}\Z'$$
where \hbox{$\Z'\onto\M$} is a vector bundle.\footnote{
Most notably, the inclusion is proper in the case of the electron
and positron bundles (\sref{sss:Dirac field}).} 
For each \hbox{$p\in\Pm$}\,, the fiber's algebraic structure
determines a projection \hbox{$\Pi_{\sst\Z}(p):\Z'\onto\Z_p$}\,,
which can be expressed as
$$\Pi_{\sst\Z}(p)=\bb_\a(p)\tn\bb^\a(p)$$
in a suitable frame adapted to $\Z_p$\,;
if \hbox{$\Z=\Pm\cart{\M}\Z'$}
then $\Pi_{\sst\Z}(p)$ is just the identity.
Analogously, the map
$$\Pi_{\sst\Zc{}^*}(p)=\bar\bb_\a(p)\tn\bar\bb^\a(p)$$
is either the identity or the projection onto \hbox{$\Zc{}_p^*\cong\Z_p$}\,.
Now, by composing the second tensor factors
with absorption and emission operators,
and doing transpositions for formal purposes,
we obtain the maps
\begin{align*}
&\Phi^+:\Pm\to\OC\tn\Z:p\mapsto\abs{a}^\a(p)\tn\bb_\a(p)~,
\\[6pt]
&\Phi^-:\Pm\to\OC\tn\Zc{}^*:p\mapsto\tra{a}{}^\a(p)\tn\bar\bb_\a(p)~.
\end{align*}

Working with a chosen observer
we label momenta \hbox{$p\in\Pm$} by their ``spatial'' part  \hbox{$p_\sbo$}
(\sref{ss:Distributional bundles and generalized frames}).
We may then select an orthonormal frame $\bigl(\bb_\a(0)\bigr)$
corresponding to \hbox{$p_\sbo=0$}\,.
In the situations of our interest one finds that there is,
for each \hbox{$p_\sbo\in\P_{\!\!\sbot}$}\,, a natural and essentially unique
unitary transformation \hbox{$K(p_\sbo):\Z_0\to\Z_{p_\sbo}$}\,,
which yields the orthonormal frames
$$\bigl(\bb_\a(p_\sbo)\bigr)\equiv\bigl(K(p_\sbo)\bb_\a(0)\bigr)=
\bigl(K\Ii\b\a(p_\sbo)\,\bb_\b(0)\bigr)~,\quad p_\sbo\in\Pm~.$$
We then note that, because of unitarity, the conjugate frames 
$\bigl(\bar\bb_\a(p_\sbo)\bigr)$ transform with the same rule,
while both the dual frame $\bigl(\bb^\a(p_\sbo)\bigr)$
and the anti-dual frame $\bigl(\bar\bb^\a(p_\sbo)\bigr)$
transform according to the inverse matrix
$\smash{\bigl(\cev K\Ii\b\a(p_\sbo)\bigr)}$\,.
We now express $\Phi^+$ and $\Phi^-$ as
\begin{align*}
&\Phi^+(p_\sbo)=\Phi^{{+}\a}(p_\sbo)\tn\bb_\a(0)\equiv
\bigl(K\Ii\a\b(p_\sbo)\,\abs{a}^\b(p_\sbo)\bigr)\tn\bb_\a(0)~,
\\[6pt]
&\Phi^-(p_\sbo)=\Phi^{{-}\a}(p_\sbo)\tn\bb_\a(0)\equiv
\bigl(K\Ii\a\b(p_\sbo)\,\tra{a}{}^\b(p_\sbo)\bigr)\tn\bb_\a(0)~.
\end{align*}

The above components $\Phi^{{+}\a}$ and $\Phi^{{-}\a}$
are written in the frame $\bigl(\bb_\a(0)\bigr)$\,,
which is independent of momentum.
Next we consider again the setting described in~\sref{ss:Quantum configuration space},
and realize that for all \hbox{$t\in\T$} we can perform spatial
Fourier transforms and anti-transforms of $\Phi^{{+}\a}$ and $\Phi^{{-}\a}$
obtaining $\OC$-valued distributions on \hbox{$(\TO\M)_t^{\sbot}$}\,.
We then get the generalized map
$$\phi=\phi^\a\,\bb_\a(0):(\TO\M)_\Tt^{\sbot}\to\OC\tn\Z$$
whose components have the expression\footnote{
The factor $(2\,p_0)^{-1/2}$
is related to the Leray form of the mass shell.} 
$$\phi^\a(x)\equiv
\frac1{(2\pi)^{3/2}}\int\frac{\dO^3 p}{\sqrt{2\,p_0}}\,
K^\a_\b(p_\sbo)\bigl(\eO^{-\iO\,\bang{p,x}}\,\abs{a}^\b(p_\sbo)
+\eO^{\iO\,\bang{p,x}}\,\tra{a}{}^\b(p_\sbo)\bigr)~,\quad
p_0\equiv(m^2+p_\sbo^2)^{1/2}~,$$
and are easily seen to fulfil the Klein-Gordon equation.
We remark that the above \emph{free field} can always be seen as a combination
of \emph{particle absorption and anti-particle emission},
where the terms ``particle'' and ``antiparticle'' refer to
the internal bundles $\Z$ and $\Zc$;
if $\Z$ is real then these coincide, and we lose the distinction.

\smallbreak

In comparison with the notion of a field defined on $\M$,
the above scheme can be seen as yielding a kind of \emph{linearized} construction.
If $(\M,g)$ is Minkovski spacetime and we have an \emph{inertial}
orthogonal decomposition \hbox{$\M=\T\,{\times}\,\X$},
then by obvious identifications we also obtain a true field over $\M$;
but note that the affine space $\X$
(the space of ``positions'' of the chosen observer)
has here a distinguished point, namely the detector's position,
so that it can be identified with a vector space.
In curved spacetime a section of a vector bundle over $\M$ can be obtained too,
but possibly only in a neighbourhood of the detector $\T$.
Without entering details,
the construction uses the local isomorphism (\sref{ss:Quantum configuration space})
of a neighbourhood of $t$ in $\M\!_t$
with a neighbourhood of $0$ in \hbox{$(\TO\M)_t^{\sbot}$}\,,
together with  parallel transport of $\bb_\a(0)$
along the spacelike geodesic from $t$ to \hbox{$x\in\M\!_t$}
relatively to a fixed background connection
(possibly related to gauge-fixing).
We stress that the components $\phi^\a(x)$ are valued in a \emph{fixed} algebra
of linear operators on the space $\QC$ of quantum states.
This is actually the extended operator space $\OC^\bullet$
(\sref{ss:Multi-particle states and elementary operators}),
but we'll indicate it as $\OC$ for notational simplicity.

The above constructions yield a so-called \emph{free quantum field},
which is an essentially unique, well-defined object,
fulfilling the Klein-Gordon equation
and determined by the underlying classical geometry.
More generally we can consider arbitrary generalised sections
\hbox{$\M\to\OC\ten{\M}\Z$}\,.
The quantum fields of a theory can be described as generalised sections
\hbox{$\M\to\EC\equiv\OC\tn\E$},
where \hbox{$\E\onto\M$} is the classical ``configuration bundle''
(this is the finite-dimensional vector bundle
whose sections are the ``pre-quantum'' fields)
and \hbox{$\EC\onto\M$} is the corresponding ``quantum bundle''.

\smallbreak
For simplicity of notation and exposition,
in the rest of this paper we'll work in flat spacetime
with a given inertial decomposition \hbox{$\M=\T\,{\times}\,\X$}
(and $\X$ is identified with a vector space as remarked above),
but we stress that most constructions and results, with proper caveats,
can be recast in a more general scenario.

\subsection{Conjugate fields}
\label{ss:Conjugate fields}

The scheme sketched in \sref{ss:Free quantum fields}
is suitable for describing bosonic \emph{and} fermionic free quantum fields,
as the differences between these two cases are dealt with by
the super-commutation rules among absorption and emission operators.
However there is a complication, related to conjugate fields,
which deserves a thorough discussion.

We begin by clarifying a notational issue.
In the standard theoretical physics literature,
complex conjugation is usually indicated by an asterisk.
In mathematics, complex conjugation is usually indicated by an overbar,
while an asterisk labels transposition.
We'll stick to the mathematics usage,\footnote{
One finds further notational variations, however.
For example in~\cite{DeGe} an asterisk stands for Hermitian transposition
while dual spaces are labeled by the symbol~$\#$.} 
and note that there is one situation of apparent conflict:
the ``Dirac adjoint'' $\bar\psi$ of a Dirac spinor $\psi$\,.
Actually it turns out that this is easily adjusted,
as the space $\W$ of 4-spinors has a natural Hermitian structure of signature $(2,2)$\,,
and \hbox{$\bar\psi\in\W^*$} is exactly the element corresponding to
the complex conjugate of $\psi$ via the induced isomorphism
\hbox{$\Wc\leftrightarrow\W^*$}.
In general,
no issue arises about denoting the Hermitian transpose of $\bar\phi$ as $\phi^\dag$,
though that is somewhat pleonastic
as one could just write $\bar\phi$ implying the isomorphism
\hbox{$\Zc\leftrightarrow\Z^*$} determined by Hermitian structure.
But note that \hbox{$\psi^\dag\equiv\bar\psi\,\g^0$},
in the Dirac context,
is the transpose of $\bar\psi$ with respect to a different,
\emph{positive definite} Hermitian structure
which is associated with the chosen observer.\footnote{
These issues were thoroughly examined in previous papers~\cite{C98,C00b,C07}.}

The above considerations are valid in the classical field context,
but in the quantum context there are further complications.
With regard to conjugation we have two possible constructions:
we can take the complex conjugate of $\phi$\,,
and also make the same construction used for $\phi$
but replacing the internal bundle $\Z$
with the ``anti-particle'' bundle $\Zc$.
Moreover we can apply transposition
in the operator algebra $\OC$, indicated by an asterisk.
In order to avoid possible confusions we'll indicate the complex conjugate
of $\phi$ by $\Ccal\phi$\,,
and reserve the symbol $\bar\phi$ for a different construction
involving the internal bundle $\Zc$.
Now in connection with the free field \hbox{$\phi=\phi^\a\,\bb_\a(0)$}
introduced in~\sref{ss:Free quantum fields} we also obtain the fields
$$\phi^*=\phi^{\a{*}}\,\bb_\a(0)~,\qquad
\Ccal\phi=\Ccal\phi_\a\,\bar\bb^\a(0)~,\qquad
\Ccal\phi^*=(\Ccal\phi_\a)^*\,\bar\bb^\a(0)~,$$
where\footnote{
We used \hbox{$K^\dag=\cev K$}.} 
\begin{align*}
\phi^{\a{*}}(x)&=\frac1{(2\pi)^{3/2}}\int\frac{\dO^3 p}{\sqrt{2\,p_0}}\,
K^\a_\b(p_\sbo)\bigl(
\eO^{-\iO\,\bang{p,x}}\,\tra{a}{}^\b(p_\sbo)
+\eO^{\iO\,\bang{p,x}}\,\abs{a}^\b(p_\sbo)\bigr)~,
\displaybreak[2]\\[8pt]
\Ccal\phi_\a(x)&=\frac1{(2\pi)^{3/2}}\int\frac{\dO^3 p}{\sqrt{2\,p_0}}\,
\cev K{}^\b_\a\,(p_\sbo)\bigl(
\eO^{\iO\,\bang{p,x}}\,\abs{a}_\b(p_\sbo)
+\eO^{-\iO\,\bang{p,x}}\,\tra{a}_\b(p_\sbo)\bigr)~,
\displaybreak[2]\\[8pt]
\Ccal\phi_\a^*(x)&=\frac1{(2\pi)^{3/2}}\int\frac{\dO^3 p}{\sqrt{2\,p_0}}\,
\cev K{}^\b_\a\,(p_\sbo)\bigl(
\eO^{\iO\,\bang{p,x}}\,\tra{a}_\b(p_\sbo)
+\eO^{-\iO\,\bang{p,x}}\,\abs{a}_\b(p_\sbo)\bigr)~.
\end{align*}

We then see that $\bar\phi{}^*$ is exactly the ``free field of the conjugate bundle'',
namely it can be obtained by the same construction as $\phi$\,,
after replacing $\Z$ with $\Zc$,
as a combination of \emph{anti-particle absorption}
and \emph{particle emission} operators.
Up to identifications which seem obvious in the matrix formalism,
$\Ccal\phi^*$ is essentially the field which in a generic setting
is usually denoted as $\phi^\dag$.

In a classical field theory one deals with fields and their complex conjugates,
which are to be replaced with $\OC$-valued fields upon quantization.
Free fields play a specially important role as
basic ``building blocks'' of field dynamics.
When we evaluate any functional of the fields in terms of free fields,
we are to replace the classical components
$\phi^\a(x)$ with the expression written in~\sref{ss:Free quantum fields}.
Which is the correct replacement for $\Ccal\phi_\a(x)$?
The answer depends on certain properties that field super-commutators must obey.
We claim that it is
$$\bar\phi_\a(x)=
\frac1{(2\pi)^{3/2}}\int\frac{\dO^3 p}{\sqrt{2\,p_0}}\,
\cev K{}^\b_\a\,(p_\sbo)\Bigl( \pm
\eO^{-\iO\,\bang{p,x}}\,\abs{a}_\b(p_\sbo)
+\eO^{\iO\,\bang{p,x}}\,\tra{a}_\b(p_\sbo) \Bigr)~,$$
where the upper sign in the first term in the integrand
holds for boson fields, while the lower sign holds for fermion fields.
Thus $\bar\phi_\a$ coincides with $\Ccal\phi_\a^*$ for bosons
but \emph{not} for fermions.
This seems to be in contrast with standard presentations,
as far as fermion fields are concerned;
however we'll argue that the minus sign is actually present in the
usual expression for the Dirac-adjoint quantum field $\bar\psi$\,,
though somewhat hidden in the intricacies of the matrix formalism.
Moreover we'll check that the required identities,
and the free-field expressions of the most important functionals,
do follow from the above prescription.\footnote{
The notion of anti-particle is often introduced,
according to an hystorical presentation,
in the discussion of the Dirac spinor field,
in relation to the \emph{equal-time commutation rules}
which quantum fields and their conjugates are required to obey
as an implementation of the principle of correspondence.
The same discussion is also offered as a justification for the introduction
of anti-commuting absorption and emission operators.} 

\remark~Because of the isomorphism \hbox{$\Zc\leftrightarrow\Z^*$},
we can equivalently view $\Z^*$ as the internal anti-particle bundle.
Both views require the Hermitian structure unless we deal with real bundles
(which by the way is exactly the case of the ghost and anti-ghost fields,
see~\sref{sss:Ghost and anti-ghost fields}).

\subsection{Recalls about propagators}
\label{ss:Recalls about propagators}

In terms of the decomposition \hbox{$\M=\T\times\X$} discussed
in~\sref{ss:Free quantum fields} we write \hbox{$x\equiv(t,x_\sbo)\in\M$}.
We also have (\sref{ss:Distributional bundles and generalized frames})
the splitting \hbox{$\TS\M\equiv\P=\P_{\!\!\spar}\oplus\P_{\!\!\sbot}$}\,.
This is the bundle of momenta, which is trivial in the flat case.
We write \hbox{$p\equiv(p_0,p_\sbo)\in\P$}.
If \hbox{$p\in\Pm\subset\P$} then
$$p_0=\Eo_m(p_\sbo)\equiv\sqrt{m^2+|p_\sbo^2|}~.$$

The evaluation of field super-commutators yields the integrals
$$\Dcal^\pm(x)\equiv\frac{\pm1}{(2\pi)^{3}}
\int\frac{\dO^3p_\sbo}{2\,p_0}\,\eO^{\mp\iO\,\bang{p,x}}~,\quad
p\in\Pm~,$$
which are well-defined distributions.
The convention of using the symbol $p_0$
as a positive ``on-shell'' function of $p_\sbo$ is common and here we'll use it,
though it could be confusing if one aims at a systematical understanding
of the relations among special generalized densities and propagators.\footnote{
In the physics literature one tries to avoid such issues,
possibly by \emph{ad hoc} spatial variable changes.} 
Despite appearance, these are full Fourier transforms,
since they can also be written in the form
$$\Dcal^\pm(x)=\frac{1}{(2\pi)^{3}}\int\frac{\dO^4p}{2\,p_0}\,
\eO^{\mp\iO\,\bang{p,x}}\,\d\bigl(p_0\mp\Eo_m(p)\bigr)~.$$
As such, they are recognized as the Fourier transform
and minus the Fourier anti-transforms of the Leray density  $\om_m$ of $\Pm$
divided by $2\pi$ and seen as a generalized density on $\P$
(\sref{ss:Distributional bundles and generalized frames}).

We'll be also involved with the partial derivatives
$$\Dcal^\pm_{,\l}(x)\equiv\tfrac{\de}{\de x^\l}\Dcal(x)=
\frac{-\iO}{(2\pi)^{3}}\int\frac{\dO^3p_\sbo}{2\,\Eo_m(p)}\,p_\l\,
\eO^{\mp\iO\,\bang{p,x}}~,\quad \l=0,1,2,3.$$
Moreover we set \hbox{$\Dcal\equiv\Dcal^+ + \Dcal^-$}, and find
$$\Dcal^+(-x)=-\Dcal^-(x)\qRq
\Dcal(-x)=-\Dcal(x)~.$$
Finally we obtain the ``zero-time'' relations
\begin{align*}
&\Dcal^+(0,x_\sbo)=-\Dcal^-(0,x_\sbo) \qRq \Dcal(0,x_\sbo)=0~,
\\[6pt]
&\Dcal^\pm_{,0}(0,x_\sbo)=-\ih\,\d(x_\sbo) \qRq
-\iO\,\d(x_\sbo)=\Dcal_{,0}(0,x_\sbo)\equiv
\Dcal^+_{,0}(0,x_\sbo)+\Dcal^-_{,0}(0,x_\sbo)~.
\end{align*}

While the generalized density $\om_m$ is observer-dependent,
the combination \hbox{$\Dcal\equiv\Dcal^+ + \Dcal^-$}
turns out to be a geometrically well-defined object,
as it is the Fourier transform of the observer-independent
Leray form\footnote{
Usually denoted as \hbox{$\d(p^2-m^2)$}\,.} 
$\om[\underline g-m^2]$ where \hbox{$\underline g(p)\equiv p^2$}.
Hence the above identities imply that $\Dcal$ and its derivatives
vanish outside the causal cone.

\subsection{Field super-commutators}
\label{ss:Field super-commutators}

The basic super-commutation rules of emission and absorption operators
(\sref{ss:Conjugation and the role of Hermitian structure})
can be rewritten in the present context as follows.
We have
$$\suc{\abs{a}^\a(p_\sbo)\,,\,\tra{a}_\b(q_\sbo)}=
\suc{\abs{a}_\b(p_\sbo)\,,\,\tra{a}{}^\a(q_\sbo)}=
\d^\a_\b\,\d(p_\sbo-q_\sbo)~,$$
while other super-commutators vanish, namely
\begin{align*}
0&=\suc{\abs{a}^\a(p_\sbo)\,,\,\abs{a}^\b(q_\sbo)}=
\suc{\tra{a}{}^\a(p_\sbo)\,,\,\tra{a}{}^\b(q_\sbo)}=
\suc{\abs{a}_\a(p_\sbo)\,,\,\abs{a}_\b(q_\sbo)}=
\suc{\tra{a}_\a(p_\sbo)\,,\,\tra{a}_\b(q_\sbo)}={}
\\[6pt]
&=\suc{\abs{a}^\a(p_\sbo)\,,\,\abs{a}_\b(q_\sbo)}=
\suc{\tra{a}{}^\a(p_\sbo)\,,\,\tra{a}_\b(q_\sbo)}=
\suc{\abs{a}_\a(p_\sbo)\,,\,\tra{a}_\b(q_\sbo)}=
\suc{\abs{a}^\a(p_\sbo)\,,\,\tra{a}{}^\b(q_\sbo)}~.
\end{align*}

Then, for any two events \hbox{$x,x'\in\M$}
we have the vanishing super-commutators\footnote{
We are writing these identities in the hypothesis that the map $\Pi_{\sst\Z}$
introduced in~\sref{ss:Free quantum fields} is the identity of the fibers
of the internal bundle $\Z$.
The main case in which it is not, namely that of Dirac fields,
will be worked out separately (\sref{sss:Dirac field})} 
\begin{align*}
&\Suc{\phi^\a(x)\,,\,\phi^\b(x')}=\Suc{\phi^\a(x)\,,\,\phi^{\b{*}}(x')}=
\Suc{\phi^{\a{*}}(x)\,,\,\phi^{\b{*}}(x')}=0~,
\\[6pt]
&\Suc{\phi^\a(x)\,,\,\phi^\b_{,\l}(x')}=
\Suc{\phi^\a(x)\,,\,\phi^{\b{*}}_{,\l}(x')}=
\Suc{\phi^{\a{*}}(x)\,,\,\phi^{\b{*}}_{,\l}(x')}=0~,
\end{align*}
Moreover we find the super-commutators
\begin{align*}
&\Suc{\phi^\a(x)\,,\,\Ccal\phi_\b(x')}=
\d^\a_\b\,\bigl(\Dcal^+(x+x')\pm\Dcal^-(x+x')\bigr)~,
\\[6pt]
&\Suc{\phi^\a(x)\,,\,\Ccal\phi^*\b(x')}=
\d^\a_\b\,\bigl(\Dcal^+(x-x')\pm\Dcal^-(x-x')\bigr)~,
\\[6pt]
&\Suc{\phi^\a(x)\,,\,\bar\phi_\b(x')}=
\d^\a_\b\,\bigl(\Dcal^+(x-x')+\Dcal^-(x-x')\bigr)\equiv
\d^\a_\b\,\Dcal(x-x')~,
\displaybreak[2]\\[6pt]
&\Suc{\phi^\a(x)\,,\,\Ccal\phi_{\b,\l}(x')}=
\d^\a_\b\,\bigl(\Dcal^+_{\!,\l}(x+x')\pm\Dcal^-_{\!,\l}(x+x')\bigr)~,
\\[6pt]
&\Suc{\phi^\a(x)\,,\,\Ccal\phi^*_{\b,\l}(x')}=
\d^\a_\b\,\bigl(-\Dcal^+_{\!,\l}(x-x')\mp\Dcal^-_{\!,\l}(x-x')\bigr)~,
\\[6pt]
&\Suc{\phi^\a(x)\,,\,\bar\phi_{\b,\l}(x')}=
-\d^\a_\b\,\bigl(\Dcal^+_{\!,\l}(x-x')+\Dcal^-_{\!,\l}(x-x')\bigr)\equiv
-\d^\a_\b\,\Dcal_{\!,\l}(x-x')~,
\\[6pt]
&\Suc{\phi^\a_{,\l}(x)\,,\,\bar\phi_\b(x')}=
\d^\a_\b\,\bigl(\Dcal^+_{\!,\l}(x-x')+\Dcal^-_{\!,\l}(x-x')\bigr)\equiv
\d^\a_\b\,\Dcal_{\!,\l}(x-x')~,
\end{align*}
where \hbox{$\bar\phi_{\a,\l}\equiv\de\bar\phi_\a/\de x^\l$} and the like,
and double signs apply to the alternative boson/fermion.

We now observe that, out of the above non-vanishing super-commutators,
those which involve $\bar\phi$ and its derivatives depend on the difference
$x\,{-}\,x'$ and are expressed in terms of the observer-independent
distribution $\Dcal$.
This fact endorses our prescription of $\bar\phi$
as the right free quantum field replacement for a classical field $\Ccal\phi$\,.
In the bosonic case $\bar\phi$ coincides with $\Ccal\phi^*$,
which in a generic context is usually indicated as $\phi^\dag$
(\sref{ss:Conjugate fields}).

At equal times ($x^0=x'^0$) we obtain
\begin{align*}
&\Suc{\phi^\a(x)\,,\,\bar\phi_\b(x')}=0~,
\\[6pt]
&\Suc{\phi^\a(x)\,,\,\bar\phi_{\b,0}(x')}=
-\Suc{\phi^\a_{,0}(x)\,,\,\bar\phi_\b(x')}=
-\iO\,\d^\a_\b\,\d(x_\sbo\,{-}\,x'_\sbo)~.
\end{align*}

\subsection{Conjugate momenta and the Hamiltonian}
\label{ss:Conjugate momenta and the Hamiltonian}

In the context of Lagrangian field theory one sets
\hbox{$\p_\a:=\de\ell/\de\phi^\a_{,0}$}
where $\ell\,\dO^4x$ is the total Lagrangian density.
In a Hamiltonian setting,
$\p_\a$ plays the role of the ``conjugate momentum'' associated with $\phi^\a$.
The required equal-time super-commutation rules are of the type
\begin{align*}
&\Suc{\phi^\a(x)\,,\,\p_\b(x')}=\pm\iO\,\d^\a_\b\,\d(x_\sbo\,{-}\,x'_\sbo)\,\rdg~,
\\[6pt]
&\Suc{\phi^\a(x)\,,\,\phi^\b(x')}=\Suc{\p_\a(x)\,,\,\p_\b(x')}=0~,
\end{align*}
with \hbox{$x\equiv(t,x_\sbo)$}\,, \hbox{$x'\equiv(t,x'_\sbo)$}\,,
\hbox{$\rdg\equiv\rdetg$}\,.
These rules are to be directly checked to hold true for free fields;
their validity for critical sections\footnote{
That is solutions of the full field equations with interactions.
} 
can then be inferred by general arguments based on the form of the dynamics.
Note that, in standard expressions written in terms of field components,
the product of field components valued at the same spacetime point
is defined by normal ordering
(\sref{ss:Multi-particle states and elementary operators}),
in order to obtain $\OC$-valued quantities.
Instead, normal ordering is \emph{not} assumed in the above rules,
which must be intended in a generalized distributional sense.

The Hamiltonian density and the Hamiltonian of a general field theory are the functionals
$$\phi\mapsto\Hcal[\phi]=\p_\a[\phi]\,\phi^\a_{,0}-\ell[\phi]~,\qquad
H[\phi](t)=\int\dO^3x_\sbo\,\Hcal[\phi](t,x_\sbo)~.$$
In particular one is interested in the \emph{free} Hamiltonian
in each sector of the theory,
obtained by dropping all interactions with other sectors
and then evaluating through free fields.
A basic example is obtained from the sector Lagrangian
$$\ell\free[\phi,\bar\phi]=\bigl(
\oh\,g^{\l\m}\,\bar\phi_{\a,\l}\,\phi^\a_{,\m}
-\oh\,m^2\,\bar\phi_\a\,\phi^\a\bigr)\,\rdg~,\quad
\l,\m=0,1,2,3,$$
whence
\begin{align*}
&\p_\a=\oh\,g^{\l0}\,\bar\phi_{\a,\l}\,\rdg~,\qquad
\p^\a=\oh\,g^{0\l}\,\phi^\a_{,\l}\,\rdg~,
\\[8pt]
&\tfrac1\rrdg\,\Hcal\!\free[\phi,\bar\phi]=\bar\phi_{\a,0}\,\phi^\a_{,0}
-\oh\,g^{\l\m}\,\bar\phi_{\a,\l}\,\phi^\a_{,\m}+\oh\,m^2\,\bar\phi_\a\,\phi^\a=
\\[6pt]
&\phantom{\tfrac1\rrdg\,\Hcal\!\free[\phi,\bar\phi]}
=\oh\,\bar\phi_{\a,0}\,\phi^\a_{,0}
-\oh\,g^{ij}\,\bar\phi_{\a,i}\,\phi^\a_{,j}+\oh\,m^2\,\bar\phi_\a\,\phi^\a~,\quad
i,j=1,2,3.
\end{align*}

By evaluation through quantum free fields we then see,
keeping the results of~\sref{ss:Field super-commutators} into account,
that the above written equal-time super-commutation rules are indeed fulfilled.
Moreover, allowing normal ordering we obtain\footnote{
We are not explicitely writing this calculation, whicht turns out
to be longer than one would expect at first sight.} 
\begin{align*}
H\!\free[\phi,\bar\phi]&=\oh\,\int\!\!\dO^3x_\sbo\,\bigl(\bar\phi_{\a,0}\,\phi^\a_{,0}
-g^{ij}\,\bar\phi_{\a,i}\,\phi^\a_{,j}+m^2\,\bar\phi_\a\,\phi^\a\bigr)(t,x_\sbo)=
\\[6pt]
&=\oh\int\!\!\dO^3p_\sbo\,p_0\,\bigl(\tra{a}{}^\b(p_\sbo)\,\abs{a}_\b(p_\sbo)
+\tra{a}_\b(p_\sbo)\,\abs{a}^\b(p_\sbo)\bigr)~,
\end{align*}
which holds for boson and fermion fields alike.

\section{Quantum fields in a gauge theory}
\label{s:Quantum fields in a gauge theory}

We now elaborate on the notion of free quantum field
in a more specialized setting,
associated with a pre-quantum gauge field theory.
In~\sref{s:Antifield formalism and BRST symmetry}, instead,
we'll explore some aspects of the complementary ``covariant'' theory,
constructed by direct replacement of the finite-dimensional 
``configuration bundle'' \hbox{$\E\onto\M$}
with the ``quantum bundle'' \hbox{$\OC\tn\E\onto\M$},
obtained via fiber tensorialization by a certain $\ZZ_2$-graded algebra $\OC$.

A preliminary remark regards the relations between mutually conjugate fields.
It is often stressed that the ghost and anti-ghost fields
are independent of each other.
Indeed, they appear asymmetrically in the Lagrangian.
By contrast, the Dirac fields $\psi$ and $\bar\psi$ are formally exchanged
by conjugation in the Lagrangian and in the field equations.
Whatever the form of the field equations, however,
any fields $\phi$ and $\bar\phi$ (\sref{ss:Conjugate fields})
could be seen as mutually independent even if they admit conjugate solutions.
This point of view is streghtened by the observation that
one obtains the field equations by varying them independently,
as well as from other considerations such as the derivation
of the free Hamiltonian (\sref{ss:Some special functionals}).

\subsection{Remarks about fiber endomorphisms}
\label{ss:Remarks about fiber endomorphisms}

For historical and convenience reasons,
the standard description of a gauge field theory exploits the notion
of a fixed \emph{structure group},
via principal bundles and vector bundles associated to them.
A complementary, equivalent view can be expressed
in terms of vector bundles smoothly endowed with some fiber structure.
The group bundle of all fiber automorphisms preserving that structure
can be non trivial.
Locally, the choice of a special frame determines a trivialization
of the group bundle and isomorphisms of the fibers to a group of matrices,
so that the structure group arises as the group of transformations
among special frames.

The basic examples in physics are the tangent bundle \hbox{$\TO\M\onto\M$}
of the spacetime manifold,
whose fibers are endowed with a Lorentzian structure,
and a complex vector bundle \hbox{$\F\onto\M$},
whose fibers represent the internal particle structure and, tipically,
are endowed with a Hermitian structure.

The bundle of all linear fiber endomorphisms of $\F$
is \hbox{$\End\!\F\cong\F\ten{\M}\F^*\onto\M$}.
The ordinary commutator makes it a \emph{Lie algebra bundle}.
Seen as a real vector bundle of fiber dimension $2n^2$
(where $n$ is the complex fiber dimension of $\F$),
it is endowed with the  distinguished real symmetric bilinear form
$$\GO:\End\!\F\times\End\!\F\to\RR:(X,Y)\mapsto\Re\Tr(X\comp Y)~,$$
whose signature (see below) turns out to be $(n^2,n^2)$.

When a Hermitian structure on $\F$ is assigned,
one also obtains the Hermitian structure on $\End\!\F$ given by
$$\HO:\End\!\F\times\End\!\F\to\CC:(X,Y)\mapsto\Tr(X^\dag\comp Y)~.$$
Moreover every endomorphism can be uniquely written
as the sum of a anti-Hermitian and a Hermitian endomorphism,
namely one obtains\footnote{
Then $\Lie$ is the Lie-algebra bundle of the group bundle
of all unitary fiber automorphism.
More generally, a fiber's symmetry may be described by a different
group bundle and its derived Lie-algebra bundle.} 
the real splitting \hbox{$\End\!\F=\Lie\oplus\iO\,\Lie$}.
Now it's easy to check that the restrictions of $\HO$
to these real $n^2$-dimensional subbundles are real Euclidean
(\ie\ positive) scalar products.
The above statement about the signature of
the real 2-form $\GO$ then follows from the observation
that $\Lie$ and $\iO\,\Lie$ are respectively characterized
by the properties \hbox{$X^\dag=-X$} and \hbox{$X^\dag=X$} for any element $X$.
Note how the assignment of a Hermitian structure on $\F$
determines a splitting of the real vector space underlying $\End\!\F$
into the direct sum of two subspaces of opposite signatures.

If $\bigl(\bb_i\bigr)$ is an orthonormal frame of $\F$
then the matrix of a section \hbox{$X:\M\to\Lie$} is anti-Hermitian.
In particular, one can always find
an orthonormal frame $\bigl(\lfr_\sI\bigr)$ of $\Lie$
related to $\bigl(\bb_i\bigr)$ by the relations
\hbox{$\lfr_\sI=\lfr\iIi\sI ij \,\bb_i\tn\bb^j$},
where the matrices $\bigl(\lfr\iIi\sI ij\bigr)$ are \emph{constant}.
Then we obtain the constant coefficients (\emph{structure constants})
$$\sco\sI\sJ\sH\equiv\bang{\lfr^\sI,[\lfr_\sJ\,,\,\lfr_\sH]}~,$$
where $\bigl(\lfr^\sI\bigr)$ is the dual frame.

\subsection{Pre-quantum fields of an essential gauge theory}
\label{ss:Pre-quantum fields of an essential gauge theory}

A gauge field theory with one fermion type can be formulated by assuming,
as the fundamental geometric data, two complex bundles over a 4-dimensional
manifold $\M$:
\smallbreak\noindent
$\bullet$~the \emph{two-spinor bundle} (or \emph{Weyl bundle}) \hbox{$\U\onto\M$}
has 2-dimensional fibers,
and the fibers of \hbox{$\weu2\U\onto\M$} are endowed with a Hermitian structure
(but \emph{not} the fibers of $\U$ itself);
\smallbreak\noindent
$\bullet$~the Hermitian bundle \hbox{$\F\onto\M$},
whose fibers describe the internal degrees of freedom of fermions besides spin.
\smallbreak

Then it turns out~\cite{C98,C00b,C07}
that that the fibers of the Hermitian subbundle
\hbox{$\H\subset\U\tn\Uc$}
are naturally endowed with a Lorentz structure,
and there is a natural Clifford morphism \hbox{$\g:\H\to\End\W$},
where \hbox{$\W:=\U\dir{\M}\Uc{}^*$} can be identified as the \emph{Dirac bundle}.
The gravitational structure is jointly described
by a \emph{scaled tetrad}\footnote{
Here $\LL$ is the space of length units.
See~\cite{C12a,JMV10} for a thorough account of unit spaces.} 
\hbox{$\Th:\TO\M\to\LL\tn\H$}
and by a linear connection $\Cs$ of \hbox{$\U\onto\M$}
(\emph{$2$-spinor connection}),
which can be included among the variables
of a comprehensive Lagrangian theory.
In this article, however, we'll assume a fixed gravitational background,
represented by an assigned couple $(\Th,\Cs)$.

Consider the following pre-quantum fields:
\smallbreak\noindent
$\bullet$~a ``matter'' field \hbox{$\psi:\M\to\W\ten{\M}\F$}\,;
\smallbreak\noindent
$\bullet$~a gauge field, namely a linear Hermitian connection of \hbox{$\F\onto\M$}.
\smallbreak

The latter can be seen as a section \hbox{$\a:\M\to\GA$},
where \hbox{$\GA\onto\M$} is an \emph{affine} bundle whose ``derived'' vector bundle
(the bundle of differences of linear Hermitian connections)
is \hbox{$\TS\M\ten{\M}\Lie\onto\M$}.
Now the quantum theory requires the fields to be sections of vector bundles,
whose fibers are tensorialized by a suitable operator algebra $\OC$
(\sref{ss:Quantum configuration space},\;\ref{ss:Free quantum fields}).
For gauge fields, this requirement is met by the choice
of a local curvature-free connection $\a_0$\,.
The field $\a$ is then represented by the difference \hbox{$A\equiv\a-\a_0$}\,.

More generally, one may consider several different $\F$ bundles
and several fermion types.
Furthermore one may consider fermion bundles like $(\FR\tn\U)\oplus(\FL\tn\Uc{}^*)$,
with different ``right'' and ``left'' components besides spin.\footnote{
We are also ignoring the Higgs field and related issues.
An account of these aspects in the pre-quantum geometric context presented here
can be found in previous papers~\cite{C10a,C14c}.} 
Here we'll limit ourselves to the essential picture,
in which however, in order to deal with the issue of the ``degrees of freedom''
of the gauge field, one also introduces
\smallbreak\noindent
$\bullet$~the \emph{ghost field} \hbox{$\gh:\M\to\Lie$}\,;
\smallbreak\noindent
$\bullet$~the \emph{anti-ghost field} \hbox{$\bar\gh:\M\to\Lie^*$}\,;
\smallbreak\noindent
$\bullet$~the \emph{Nakanishi-Lautrup field} \hbox{$n:\M\to\Lie$}\,.
\smallbreak
The latter has essentially the role of an auxiliary field,
whose components, as a consequence of the Euler-Lagrange equations,
turn out to have the expression
\hbox{$n^\sI={\smash{\frac{-1}{\xi\rrdg}}}\de_\l(g^{\l\m}\rdg A^\sI_\m)$}
where $\xi$ is a real constant.
The corresponding quantized fields $\gh$ and $\bar\gh$
will be assumed to be fermionic,
while $n$ is bosonic.
From the theory's total Lagrangian~\cite{C14b} one derives the conjugate momenta
\begin{align*}
&\p_{\a i}[\psi]=
\ih\,(\bar\psi\,\g^0)_{\a i}\,\rdg~,\qquad
\p^{\a i}[\bar\psi]=\ih\,(\g^0\,\psi)^{\a i}\,\rdg~,\qquad
\p^a_\sI[A]=\bigl(F\Ii{0\l}\sI+g^{0\l}\,n_\sI\bigr)\,\rdg~,
\\[6pt]
&\p_\sI[\gh]=g^{0\l}\,\bar\gh_{\sI,\l}\,\rdg~,\qquad
\p^\sI[\bar\gh]=-g^{0\l}\,\gh^\sI_{;\l}\,\rdg\equiv
-g^{0\l}\,(\gh^\sI_{,\l}+\sco\sI\sJ\sH\,\gh^\sJ\,A_\l^\sH)\,\rdg~,
\end{align*}
where $F$ denotes the curvature tensor of $A$,
$\a$ is a Dirac spinor index
and $i$, $\scriptstyle{I}$ are indices
in the fibers of $\F$ and $\Lie$\,, respectively.

\subsection{Gauge theory's quantum free fields}
\label{ss:Gauge theory's quantum free fields}

\subsubsection{Dirac field}
\label{sss:Dirac field}

We now inspect the case of the electron field,
namely the internal bundle is the bundle $\W$ of Dirac spinors
(these results straightforwardly carry over to $\W\tn\F$
for fermions with larger internal structure).
Then we'll briefly comment about the formal differences with usual presentations.

We first recall~\cite{C00b,C07} that the ``semi-trivial'' bundle
\hbox{$\Pm\cart{\M}\W\onto\Pm$}
has the distinguished decomposition \hbox{$\W^+\dir{\Pm}\W^-$},
where \hbox{$\W^\pm_p:=\ker(m\mp\g_p)$}\,.
The bundles \hbox{$\W^\pm\onto\Pm$} are mutually orthogonal in the Hermitian metric
associated with Dirac conjugation,
which has the signature $({+}\,{+}\,{-}\,{-})$\,;
the sign of its restriction to $\W^\pm$ is the same as the label.
A \emph{Dirac frame}
$$\bigl(\ze_\a(p)\bigr)\equiv\bigl(\uu_\sA(p)\,;\,\vv_\sB(p)\bigr)~,\quad
\a=1,2,3,4~,~~{\scriptstyle A},{\scriptstyle B}=1,2~,$$
is adapted to the above decomposition at \hbox{$p\in\Pm$}\,.
We have a distinguished transformation \hbox{$K(p_\sbo):\W\to\W$}
expressing it in terms of a frame independent of $p$,
\eg\ the Dirac frame $\bigl(\ze_\a(0)\bigr)$
associated with the chosen observer. Namely
\hbox{$\ze_\a(p_\sbo)=K\Ii\b\a(p_\sbo)\,\ze_\b(0)$}\,,
where the \hbox{$4\,{\times}\,4$} matrix of $K(p_\sbo)$
in the frame $\bigl(\ze_\a(0)\bigr)$ can be expressed as
$$K(p_\sbo)=\sqrt{\tfrac{m}{2\,(\Eo_m(p)+m)}}\,
\bigl(\id+\tfrac1m\,p_\l\,\g^\l\,\g_0\bigr)~,\qquad
(m^2+|p_\sbot|^2)^{1/2}\equiv \Eo_m(p_\sbo)\equiv p_0>0~.$$

This is essentially the transformation $K$ appearing
in the definition of the components of the free quantum field
(\sref{ss:Free quantum fields}),
but there is a slight complication:
the particle (electron) and anti-particle (positron) bundles
are now $\W^+$ and $\Wc{}^-$,
so they are not mutually conjugate bundles.
Accordingly, we introduce the absorption and emission operators
\begin{align*}
&\abs{a}^\sA(p_\sbo)\equiv\abs{a}[\Xsf^p\tn\uu^\sA(p)\bigr]~,
&&\tra{c}{}^\sA(p_\sbo)\equiv
\tra{a}[\Xsf_p\tn\bar\vv^\sA(p_\sbo)\bigr]~,
\\[6pt]
&\abs{c}_\sA(p_\sbo)\equiv
\abs{a}[\Xsf^p\tn\bar\vv_\sA(p_\sbo)\bigr]~,
&&\tra{a}_\sA(p_\sbo)\equiv\tra{a}[\Xsf_p\tn\uu_\sA(p_\sbo)\bigr]~,
\end{align*}
and obtain the non-vanishing anti-commutators
$$\bigl\{\abs{a}^\sA(p_\sbo)\,,\,\tra{a}_\sB(q_\sbo)\bigr\}=
\bigl\{\abs{c}_\sB(p_\sbo)\,,\,\tra{c}{}^\sA(q_\sbo)\bigr\}=
\d^\sA_\sB\,\d(p_\sbo-q_\sbo)~.$$
Repeating the construction seen in~\sref{ss:Free quantum fields} we now
express the elementary operators in the frame $\bigl(\ze_\a(0)\bigr)$\,, namely
\begin{align*}
&\abs{a}^\a(p)=K\Ii\a\sA(p)\,\abs{a}^\sA(p)~,
&&\tra{c}{}^\a(p)=K\Ii\a{\sA{+}2}(p)\,\tra{c}{}^\sA(p)~,
\\[6pt]
&\abs{c}_\a(p)=\cev K\Ii{\sA{+}2}\a(p)\,\abs{c}_\sA(p)~,
&&\tra{a}_\a(p)=\cev K\Ii\sA\a(p)\,\tra{a}_\sA(p)~,
\end{align*}
whence by straightforward calculations we get the anti-commutators
\begin{align*}
&\bigl\{\tra{a}_\a(p)\,,\,\abs{a}^\b(q)\bigr\}=
\tfrac1{2m}\,(m\,\id+p_\l\,\g^\l)\,\d(p_\sbo-q_\sbo)~,
\\[6pt]
&\bigl\{\abs{c}_\a(p)\,,\,\tra{c}{}^\b(q)\bigr\}=
\tfrac1{2m}\,(m\,\id-p_\l\,\g^\l)\,\d(p_\sbo-q_\sbo)~,
\\[6pt]
&\bigl\{\tra{a}_\a(p)\,,\,\tra{c}{}^\b(q)\bigr\}=
\bigl\{\abs{c}_\a(p)\,,\,\abs{a}^\b(q)\bigr\}=0~.
\end{align*}

Now, according to the general prescription introduced
in~\sref{ss:Free quantum fields},
we consider the free fields $\psi$ and $\bar\psi$ whose components
in the frame $\bigl(\ze_\a(0)\bigr)$ are
\begin{align*}
&\psi^\a(x)=\frac1{(2\pi)^{3/2}}\int\frac{\dO^3p_\sbo}{\sqrt{2\,p_0}}
\bigl(\eO^{-\iO\,p\,x}\,\abs{a}{}^\a(p_\sbo)
+\eO^{\iO\,p\,x}\,\tra{c}{}^\a\,(p_\sbo)\bigr)~,
\\[8pt]
&\bar\psi_\a(x)=\frac1{(2\pi)^{3/2}}\int\frac{\dO^3p_\sbo}{\sqrt{2\,p_0}}
\bigl(-\eO^{-\iO\,p\,x}\,\abs{c}_\a(p_\sbo)
+\eO^{\iO\,p\,x}\,\tra{a}_\a(p_\sbo)\bigr)~.
\end{align*}
It's not difficult to check that these fulfil the Dirac equation
and the conjugate Dirac equation, respectively.

\remark~As in the generic fermion case examined in~\sref{ss:Conjugate fields},
the minus sign in the expression of $\bar\psi_\a(x)$ above is needed in order to obtain
the correct supercommutator identities and expressions of field functionals
in terms of basic operators.
In order to make a thorough comparison with the matrix formulas
found in usual presentations,
we could adjust some conventions and absorb that sign
into the definition of $\abs{c}_\a(p_\sbo)$\,,
and also relate this
to the \emph{negative} Hermitian metric of the positron sector.

\subsubsection{Gauge field}
\label{sss:Gauge field}

The free gauge field is defined by
\begin{align*}
&A_\l^\sI(x)=
\frac1{(2\pi)^{3/2}}\int\frac{\dO^3 p}{\sqrt{2\,p_0}}\,\bigl(
\eO^{-\iO\,\bang{p,x}}\,\abs{b}_\l^\sI(p_\sbo)
+\eO^{\iO\,\bang{p,x}}\,\tra{b}{}_\l^\sI(p_\sbo)\bigr)~,
\\[6pt]
&\abs{b}_\l^\sI(p_\sbo)\equiv\abs{a}[\Xsf^p\tn\ee_\l\tn\lfr^\sI]~,\qquad
\tra{b}{}_\l^\sI(p_\sbo)\equiv\tra{a}[\Xsf_p\tn(\ee_\l)^\flat\tn(\lfr^\sI)^\#]~,
\end{align*}
where $\bigl(\ee_\l\bigr)$ is a possibly orthonormal spacetime frame
and $(\ee_\l)^\flat$ is the co-vector frame
associated to it via the spacetime metric;
$\bigl((\lfr^\sI)^\#\bigr)$ is the frame of $\Lie$ associated to the frame
$\bigl(\lfr^\sI\bigr)$ of $\Lie^*$
via the metric $\GO$ (\sref{ss:Remarks about fiber endomorphisms}).
Here we are not discussing frames adapted to gauge symmetry.

\subsubsection{Ghost and anti-ghost fields}
\label{sss:Ghost and anti-ghost fields}

The ghost and anti-ghost fields
(\sref{ss:Pre-quantum fields of an essential gauge theory})
are distinct, independent fields,
the isomorphism \hbox{$\Lie\cong\Lie^*$} notwithstanding.
One could view the couple $(\gh,\bar\gh)$ as a unique field
\hbox{$\M\to\End\!\F\equiv\Lie\oplus\iO\,\Lie$}\,,
namely a section of the complexified bundle of \hbox{$\Lie\onto\M$}.
The situation is then somewhat similar to that of the Dirac field,
but simpler as we do not have to deal with frames dependent on momenta.
Also note that seeing mutually conjugate fields as valued in
mutually dual internal bundles is indeed consistent with a general view,
valid both in the real case and in the complex case with a Hermitian structure
(see the remark concluding~\sref{ss:Conjugate fields}).

According to the scheme presented in~\sref{ss:Free quantum fields},
the quantum free fields $\gh$ and $\bar\gh$ are defined to have the components
\begin{align*}
&\gh^\sI(x)=
\frac1{(2\pi)^{3/2}}\int\frac{\dO^3 p_\sbo}{\sqrt{2\,p_0}}\,\bigl(
\eO^{-\iO\,\bang{p,x}}\,\abs{g}^\sI(p_\sbo)
+\eO^{\iO\,\bang{p,x}}\,\tra{k}{}^\sI(p_\sbo)\bigr)~,
\\[6pt]
&\bar\gh_\sI(x)=
\frac1{(2\pi)^{3/2}}\int\frac{\dO^3 p_\sbo}{\sqrt{2\,p_0}}\,\bigl(
-\eO^{-\iO\,\bang{p,x}}\,\abs{k}_\sI(p_\sbo)
+\eO^{\iO\,\bang{p,x}}\,\tra{g}_\sI(p_\sbo)\bigr)~,
\end{align*}
where \hbox{$p_0=|p_\sbo|$} (\hbox{$m=0$}) and
\begin{align*}
&\abs{g}^{\sI}(p_\sbo):=\abs{a}[\Xsf^{p}\tn\lfr^\sI]~,\quad
\tra{g}_\sI(p_\sbo):=\tra{a}[\Xsf_p\tn\lfr_\sI]~,
\\[6pt]
&\abs{k}_\sI(p_\sbo):=\abs{a}[\Xsf^{p}\tn\lfr_\sI]~,\quad
\tra{k}{}^\sI(p_\sbo):=\tra{a}[\Xsf_p\tn\lfr^\sI]~.
\end{align*}
The minus sign in $\bar\gh_\sI$ is related to the fact
that these are assumed to be fermion fields.

\subsection{Some special functionals}
\label{ss:Some special functionals}

Certain density functionals of the fields have special roles,
and their spatially integrated evaluation through free fields
yields remarkably simple expressions, independent of time.
We are not going to write detailed calculations,
but we stress that the following reported results
for the fermion fields critically depend on our assumptions
on the form of the free conjugate fields.
In particular, note the expression of free $4$-momentum for ghosts.

All expressions are written by allowing normal ordering.

\subsubsection*{Dirac charge}
The Dirac current is the $3$-form $\bang{\bar\psi\g^\l\psi}\,\dO x_\l$\,,
where \hbox{$\dO x_\l\equiv\de x_\l|\dO^4x$}\,.
Its restriction to constant-time hypersurfaces is the scalar density
\hbox{$\bang{\bar\psi\g^0\psi}\,\dO x_0\equiv\bang{\bar\psi\g^0\psi}\,\dO^3x_\sbo$}\,,
whose global value is the Dirac charge
$$Q\spec{Dir}\equiv\int\!\!\dO^3x_\sbo\,\bar\psi(x)\g^\l\psi(x)=
\int\frac{\dO^3p_\sbo}{2\,m}\,\bigl(\tra{a}_\sA(p_\sbo)\,\abs{a}^\sA(p_\sbo)
-\tra{c}{}^\sA(p_\sbo)\,\abs{c}_\sA(p_\sbo)\bigr)~.$$

\subsubsection*{Dirac momentum}
We recall that in flat spacetime\footnote{
See~\cite{CM} for an extension to curved spacetimes.} 
the canonical energy-momentum tensor evaluated through a field $\phi$ is a section
\hbox{$\Tcal[\phi]:\M\to\TS\M\tn\weu3\TS\M$},
with the coordinate expression
$$\Tcal[\phi]=
(\phi^\a_{,\l}\,\de^\m_\a\ell[\phi]-\ell[\phi]\,\d^\m_\l)\,\dO x^\l\tn\dO x_\m~.$$
In the case of the free Dirac field we obtain
$$\Tcal[\psi,\bar\psi]=
\ih\,\bigl(-\bar\psi_{,\l}\,\g^\m\,\psi+\bar\psi\,\g^\m\,\psi_{,\l}\bigr)\,
\dO x^\l\tn\dO x_\m~.$$

The corresponding \emph{4-momentum density}
$$\Pcal\!_\l\,\dO x^\l\tn\dO x_0=
\ih\,\bigl(-\bar\psi_{,\l}\,\g^0\,\psi+\bar\psi\,\g^0\,\psi_{,\l}\bigr)\,
\dO x^\l\tn\dO x_0$$
can be defined by a suitable pull-back via the inclusions
of the constant-time hyper-planes into $\M$,
and \emph{4-momentum} is defined to be the 1-form $P_\l\,\dO x^\l$. We obtain
\begin{align*}
P_\l&\equiv \ih \int\dO x_0\,\bigl(
-\bar\psi_{,\l}\,\g^0\,\psi+\bar\psi\,\g^0\,\psi_{,\l}\bigr)\equiv
\ih \int\dO^3x_\sbo\,\bigl(
-\bar\psi_{,\l}\,\g^0\,\psi+\bar\psi\,\g^0\,\psi_{,\l}\bigr)=
\\[6pt]
&=\int\frac{\dO^3p_\sbo}{2\,m}\,p_\l \bigl(
\tra{a}_\sA(p_\sbo)\,\abs{a}^\sA(p_\sbo)
+\tra{c}{}^\sA(p_\sbo)\,\abs{c}_\sA(p_\sbo)\bigr)~.
\end{align*}

\subsubsection*{Dirac Hamiltonian}
The Hamiltonian density $\Hcal$ of a Lagrangian field theory
has the expression
$$\Hcal[\phi]=\p_\a[\phi]\,\phi^\a_{,0}-\ell[\phi]~.$$
The \emph{free field Hamiltonian density},
for each sector of any theory,
is obtained by keeping only those terms in which
no contribution from other sectors appears,
and evaluating it through free fields.
For the Dirac sector we obtain
$$\Hcal\free[\psi,\bar\psi]=
\ih\,(\bar\psi_{,i}\,\g^i\,\psi-\bar\psi\,\g^i\,\psi_{,i})+m\,\bar\psi\,\psi=
\ih\,(\bar\psi\,\g^0\,\psi_{,0}-\bar\psi_{,0}\,\g^0\,\psi)~,$$
which is just the $0$-component of $4$-momentum
(the latter equality was written by taking the Dirac equation into account).

\subsubsection*{Ghost momentum and Hamiltonian}
In the ghost-antighost sector, components of the canonical energy-momentum tensor
have the expression
$$\Tcal{}^\m_\l[\gh,\bar\gh]=g^{\m\n}\,(\bar\gh_{\sI,\n}\,\gh^\sI_{,\l}
+\bar\gh_{\sI,\l}\,\gh^\sI_{;\n})
-g^{\n\r}\,\bar\gh_{\sI,\n}\,\gh^\sI_{;\r}\,\d^\m_\l~,$$
where \hbox{$\gh^\sI_{;\n}\equiv\na_\n\gh^\sI\equiv
\gh^\sI_{,\n}+\sco\sI\sJ\sH\,\gh^\sJ\,A_\n^\sH$}\,.
Then the components of the ghost $4$-momentum density,
evaluated through free fields, are
$$\Tcal{}^0_\l[\gh,\bar\gh]=\bar\gh_{\sI,0}\,\gh^\sI_{,\l}
+\bar\gh_{\sI,\l}\,\gh^\sI_{,0}
-g^{\n\r}\,\bar\gh_{\sI,\n}\,\gh^\sI_{,\r}\,\d^0_\l~.$$
By spatial integration we then obtain the free ghost $4$-momentum,
with components
$$P_\l[\gh,\bar\gh]=\int\!\!\dO^3p_\sbo\; p_\l \bigl(
\tra{k}{}^\sI(p_\sbo)\,\abs{k}_\sI(p_\sbo)
+\tra{g}_\sI(p_\sbo)\,\abs{g}^\sI(p_\sbo)\bigr)~,\qquad
p_0\equiv|p_\sbo|\,,~m=0\,.$$

Again, we easily check that the free Hamiltonian density
is \hbox{$\Hcal\free[\gh,\bar\gh]=\Tcal{}^0_0[\gh,\bar\gh]$}\,,
so that the $0$-component of $P$ coincides with the free Hamiltonian
\hbox{$H\!\free[\gh,\bar\gh]\equiv\int\dO^3x_\sbo\,\Hcal\free[\gh,\bar\gh]$}\,.

\subsubsection*{Faddeev-Popov current}
The one parameter transformation
\hbox{$\gh\to\eO^\t\,\gh\,,~\bar\gh\to\eO^{-\t}\,\bar\gh$}\,, \hbox{$\t\in\RR$}\,,
obviously preserves the Lagrangian.
The corresponding infinitesimal symmetry
(\sref{ss:BRST symmetry in Lagrangian field theory})
determines the \emph{Faddeev-Popov current}
\hbox{$\Jcal\!\spec{FP}=\Jcal\!\spec{FP}^{\:\l}\,\dO x_\l$} where
(in orthonormal spacetime coordinates)
$$\Jcal\!\spec{FP}^{\:\l}=
g^{\l\m}\,(\bar\gh_{\sI,\m}\,\gh^\sI-\bar\gh_\sI\,\gh^\sI_{;\m})~.$$
Evaluating $\Jcal\!\spec{FP}^{\:\l}$ through free fields,
integrating on constant-time hyperplanes
and allowing normal ordering we find
\begin{align*}
\int\!\!\dO^3x_\sbo\,\Jcal\!\spec{FP}^{\:\l}(x)&=
g^{\l\m} \int\!\!\dO^3x_\sbo\,
\bigl(\bar\gh_{\sI,\m}\,\gh^\sI-\bar\gh_\sI\,\gh^\sI_{,\m}\bigr)(x)=
\\[6pt]
&=\iO\,g^{\l\m}\int\!\!\frac{\dO^3p}{p_0}\,p_\m\,\bigl(
\tra{k}{}^\sI(p_\sbo)\,\abs{k}_\sI(p_\sbo)+
\tra{g}_\sI(p_\sbo)\,\abs{g}^\sI(p_\sbo)\bigr)~.
\end{align*}

\subsection{Canonical supercommutation rules}
\label{ss:Canonical supercommutation rules}

In order to be convinced of the consistency of our setting
we should recover the basic supercommutators among the free fields
evaluated at different events \hbox{$x,x'\in\M$}.
We already did that in the generic setting,
which also includes ghost fields and unconstrained gauge fields.
As for the Dirac field we obtain
$$\bigl\{\bar\psi_\a(x)\,,\,\psi^\b(x')\bigr\}=
\tfrac1{2m}\,\bigl((-m\,\id+\iO\,\g^\l\,\de_\l)\Dcal(x\,{-}\,x')
\bigr)\Ii\b\a~.$$
Moreover for equal-time events (\hbox{$x^0=x'^0$}) we obtain
$$\bigl\{(\bar\psi\g^0)_\a(x)\,,\,\psi^\b(x')\bigr\}=
\bigl\{\bar\psi_\a(x)\,,\,(\g^0\,\psi)^\b(x')\bigr\}=
\tfrac1{2m}\,\d\Ii\b\a\,\d(x_\sbo-x'_\sbo)~.$$

Next we want to check the equal-time super-commutation rules
between field components and conjugate momenta, in each sector,
to be of the general form written in~\sref{ss:Conjugate momenta and the Hamiltonian}.
We work in flat spacetime and set \hbox{$\rdg=1$}\,.
Considering the above identity we'd rather write
the canonical momentum conjugate to the Dirac field $\psi$ as
\hbox{$\p_\a=2\,\iO\,m\,(\bar\psi\,\g^0)_\a$}\,.
However the factor $(2m)^{-1}$ in the anti-commutator
can be absorbed by inserting factors $\sqrt{2m}$
in the definitions of $\psi$ and $\bar\psi$
(this is indeed found in the literature),
and we obtain
$$\bigl\{\p_\a(x)\,,\,\psi^\b(x')\bigr\}=
\iO\,\d\Ii\b\a\,\d(x_\sbo-x'_\sbo)~,\qquad
\p_\a=\iO\,(\bar\psi\g^0)_\a~.$$
The expression for $\p_\a$ derived from the Lagrangian
(\sref{ss:Pre-quantum fields of an essential gauge theory})
has a further factor $\oh$\,, which can be made to disappear
by changing the Lagrangian via the addition of a suitable divergence term.
Similar results hold for the conjugate sector, with
\hbox{$\p^\a=\iO\,(\g^0\psi)^\a$}.

\smallbreak
The expression
\hbox{$\p^\l_\sI=\bigl(F\Ii{0\l}\sI+g^{\l0}\,n_\sI\bigr)$}
for the canonical momentum conjugate to the gauge field
(\sref{ss:Pre-quantum fields of an essential gauge theory})
contains the term $n_\sI$ which commutes with everything,
so it may seem that it could be just dropped.\footnote{
Indices related to $\Lie$ are raised and lowered via the Euclidean metric
introduced in~\sref{ss:Remarks about fiber endomorphisms}.} 
However the remaining term $F\Ii{0\l}\sI$
does not possess the required property,
so that instead one keeps both terms and uses the replacement
\hbox{$n^\sI\to -{\smash{\frac1\xi}}\,g^{\l\m}\,A^\sI_{\l,\m}$}\,,
which is just the Euler-Lagrange field equation for $n$
(in orthonormal coordinates).
For \hbox{$\xi=1$} (the ``Feynman gauge'') we get
$$\p^\l_\sI=g^{\l\m}\,(-A_{\sI\m,0}+A_{\sI 0,\m}
-\scc_{\sI\sJ\sH}\,A_\m^\sJ\,A_0^\sH)
-g^{\l0}\,g^{\m\n}\,A^\sI_{\n,\m}~.$$
Since $A$ is a boson field, it obeys the standard commutation rules.
In particular, one easily checks
that the spatial derivatives of $A$\,, and the components of $A$ itself,
do not contribute to the equal-time commutator with $A$\,.
The part of $\p^\l_\sI$ which contains time derivatives of $A$
is just \hbox{$-g^{\l\m}\,A_{\sI\m,0}\equiv -A^\m_{\sJ,0}$}\,,
so that at equal times we eventually have
$$\bigl[A_\l^\sI(t,x_\sbo)\,,\,\p^\m_\sJ(t,x'_\sbo)\bigr]=
\bigl[A^\m_{\sJ,0}(t,x'_\sbo)\,,\,A_\l^\sI(t,x_\sbo)\bigr]=
-\iO\,\d^\m_\l\,\d^\sI_\sJ\,\d(x_\sbo-x'_\sbo)~.$$

\smallbreak
Finally we consider ghosts and anti-ghosts.
At equal times we have
\begin{align*}
&\Suc{\bar\gh_{\sJ,0}(x)\,,\,\gh^\sI(x')}=
-\Suc{\gh^\sI_{,0}(x)\,,\,\bar\gh_\sJ(x')}=
\iO\,\d^\sI_\sJ\,\d(x_\sbo-x'_\sbo)~,
\\[6pt]
&\Suc{\bar\gh_\sI(x)\,,\,\gh^\sJ(x')}=
\Suc{\bar\gh_\sI(x)\,,\,A_0^\sJ(x')}=0~,
\end{align*}
whence also
\hbox{$\Suc{\bar\gh_\sI(x)\,,\,\gh^\sJ(x') A_0^\sH(x')}=0$}\,.
From the expressions of the canonical momenta conjugate to $\gh^\sI$ and $\bar\gh_\sI$
(\sref{ss:Pre-quantum fields of an essential gauge theory}),
using the shorthand
\hbox{$\gh^\sI_{;\m}\equiv\gh^\sI_{,\m}+\sco{\sI}{\sJ\sH}\,\gh^\sJ A_\m^\sH$}\,,
we then get
\begin{align*}
&\bigl\{\gh^\sI(x)\,,\,\p_\sJ(x')\bigr\}=
\bigl\{\bar\gh_{\sJ,0}(x')\,,\,\gh^\sI(x)\bigr\}=
\iO\,\d^\sI_\sJ\,\d(x_\sbo-x'_\sbo)~,
\\[8pt]
&\bigl\{\bar\gh_\sJ(x)\,,\,\p^\sI(x')\bigr\}=
-\bigl\{\gh^\sI_{;0}(x')\,,\,\bar\gh_\sJ(x)\bigr\}=
\iO\,\d^\sJ_\sI\,\d(x_\sbo-x'_\sbo)~.
\end{align*}

\section{Antifield formalism and BRST symmetry}
\label{s:Antifield formalism and BRST symmetry}

In a previous paper~\cite{C14b} we tretated quantum fields
as sections of a \emph{quantum bundle}
obtained as a fiber tensorialization of a finite-dimensional vector bundle
by a certain infinite-dimensional $\ZZ_2$-graded algebra.
The fundamental differential geometric notions for quantum bundles,
and a related jet bundle approach to Lagrangian field theory and symmetries,
were studied there by exploiting Fr\"olicher's notion of 
smoothness~\cite{Fr,FK,KM,CJK06,KolarModugno98}.
In that context we proposed a jet bundle formulation
of Lagrangian field theory and symmetries,
and in particular of the BRST symmetry,
consistent with an ample literature treating this subject in finite-dimensional
bundles~\cite{MaMo83b,KrupkaSanders08,KrupkaKrupkovaSanders10}.
Here we resume that notion of quantum bundle, with some notational adaptations,
taking the F-smooth structure for granted.
We'll see how antifield sectors
and Batalin-Vilkovisky algebra~\cite{BV81,Fiorenza04,SchwartzA93}
naturally arise in that context,
and examine their relation to the BRST symmetry of a gauge theory
of the previously considered type.

\subsection{Quantum bundles and quantum polynomials}
\label{ss:Quantum bundles and quantum polynomials}

The classical configuration vector bundle \hbox{$\E\onto\M$}
splits as the fibered direct sum
$$\E=\E_\gradezero\dir{\M}\E_\gradeone~,$$
where the two components respectively correspond
to the bosonic and fermionic sectors (upon quantization).
The operator algebra $\OC$
introduced in~\sref{ss:Quantum configuration space} is $\ZZ_2$-graded,
so it splits as \hbox{$\OC=\OC_\gradezero\oplus\OC_\gradeone$}\,. We write
$$\OC\tn\E=\EC\dir{\M}\td\EC=
\EC_\gradezero\dir{\M}\EC_\gradeone\dir{\M}
\td\EC_\gradezero\dir{\M}\td\EC_\gradeone~,$$
where
$$\EC_\gradezero\equiv\OC_\gradezero\tn\E_\gradezero~,\qquad
\EC_\gradeone\equiv\OC_\gradeone\tn\E_\gradeone~,\qquad
\td\EC_\gradezero\equiv\OC_\gradezero\tn\E_\gradeone~,\qquad
\td\EC_\gradeone\equiv\OC_\gradeone\tn\E_\gradezero~,$$
so that quantum fields can be described as sections
\hbox{$\M\to\EC\equiv\EC_\gradezero\dir{\M}\EC_\gradeone$}\,.
Similarly we write
$$\OC\tn\E^*=\EC^*\dir{\M}\td\EC^*=
\EC^*_\gradezero\dir{\M}\EC^*_\gradeone\dir{\M}
\td\EC^*_\gradezero\dir{\M}\td\EC^*_\gradeone$$
where \hbox{$\EC^*_\gradezero\equiv\OC_\gradezero\tn\E^*_\gradezero$} and the like.
Note that an asterisk, according to these notations,
indicates duality only in a restricted sense,
though elements in $\EC^*$ can indeed be seen as linear $\OC$-valued functions.

Starting from the above constructions,
after detailing the convenient smooth structure for these bundles,
one finds that many basic notions in finite-dimensional differential geometry
are naturally extended to this quantum setting.
In particular,
tensor products and contractions in the fibers of quantum bundles over $\M$
still belong to similarly constructed quantum bundles.
Also, the notions of tangent, vertical and jet bundles
can be straightforwardly introduced,
together with the notion of a connection and various related topics.

Usually on takes \emph{linear} coordinates $\bigl(y^i\bigr)$
on the fibers of the classical bundle $\E$.
These can be viewed as $\OC$-valued coordinates on the fibers of $\EC$,
so that if \hbox{$\phi:\M\to\EC$} is a section
then we write its component expression as \hbox{$\phi=\phi^i\,\de y_i$}\,,
with \hbox{$\phi^i\equiv y^i\comp\phi:\M\to\OC$}.
We stress that the product of field components at the same spacetime point
is ``supercommutative'',
a fact that may non-trivially affect various coordinate expressions
(when compared with the corresponding classical ones).

We now set \hbox{$\E^{{*}r}\equiv\E^*\ten{\M}\mdots\ten{\M}\E^*$}
($r$ factors) and
\hbox{$\EC^{{*}r}\equiv\EC^*\ten{\M}\mdots\ten{\M}\EC^*\subset\OC\tn\E^{{*}r}$}.
A section \hbox{$f:\M\to\EC^{{*}r}$} can be written as
$$f=f_{i_1\dots i_r}\,y^{i_1}{\sst{\tn\mdots\tn}}y^{i_r}~,\qquad
f_{i_1\dots i_r}:\M\to\OC~,$$
and can be viewed as a polynomial function of degree $r$ on $\EC$,
denoted by the same symbol if no confusion arises, by writing
\hbox{$f(\phi)\equiv f(\phi,\dots,\phi)=
f_{i_1\dots i_r}\,\phi^{i_1}\mdots\phi^{i_r}$}\,.
As a rule, the classical coordinates pertain to sectors of definite parity,
so that the components of $\phi$ either commute or anti-commute
and we can also write
$$f=f_{i_1\dots i_r}\,y^{i_1}{\sst{\swe\mdots\swe}}y^{i_r}=
f_{i_1\dots i_p\,j_1\dots j_q}\,
y_\gradezero^{i_1}{\sst{\ve\mdots\ve}}y_\gradezero^{i_p}{\sst\tn}
y_\gradeone^{j_1}{\sst{\we\mdots\we}}y_\gradeone^{j_q}~,\qquad
p+q=r~,$$
where $y_\gradezero^i$ and $y_\gradeone^j$ are the coordinates
in the bosonic and fermionic sectors, respectively.
Hence, fiber polynomials of degree $r$ on $\EC$ can be represented as sections
\hbox{$\M\to\OC\tn\FC^r$} where
$$\FC^r\equiv\mathop{\textstyle\bigoplus}\limits_{p+q=r}\FC^{p,q}\equiv
\mathop{\textstyle\bigoplus}\limits_{p+q=r}
\bigl(\vee^p\E^*_\gradezero\tn\weu{q}\E^*_\gradeone\bigr)$$
(with all products fibered over $\M$).
The space
$$\FC\equiv\mathop{\textstyle\bigoplus}\limits_{r=0}^\infty\FC^r\cong
{\vee}\E^*_\gradezero\otimes{\wedge}\E^*_\gradeone$$
yields all fiber polynomials.
We stress that the exterior algebra $\wedge\E^*_\gradeone$ is also
a $\ZZ_2$-graded algebra,
so that for each $r$ we have a graded splitting
\hbox{$\FC^r=\FC^r_{\!\gradezero}\oplus\FC^r_{\!\gradeone}$}\,,
the parity of $\FC^{p,q}$ being \hbox{$q\;({\mathrm{mod}}\;2)$}\,.

We also observe that,
while a fiber polynomial on $\EC$ is an element in $\OC\tn\FC$,
considering just the space $\FC$,
which is constituted of all polynomials with numeric coefficients,
is usually sufficient,
as one starts with classical functions which are then applied to
elements in $\EC$.

While the classical linear fiber coordinates $\bigl(y^i\bigr)$
can be seen as $\OC$-valued coordinates on $\EC$,
the standard definition of partial derivative
\hbox{$\de_i f\equiv\de f/\de y^i$} doesn't work
for a quantum function \hbox{$f\in\FC$}.
However, recalling the notion of interior product in $\ZZ_2$-graded algebras,
we can introduce the \emph{left} and {right partial derivatives}
$$\lde_i f\equiv \de y_i|f~,\qquad
f\rde_i\equiv(-1)^{\grade{i}\grade{f}}\lde_if=f|\de y_i~,$$
where \hbox{$\grade{i}\equiv\grade{y^i}$} is the parity of the related sector.
These fulfill
$$\lde_i(fg)=(\lde_if)g+(-1)^{\grade{i}\grade{f}}f\,\lde_i g~,\qquad
(fg)\rde_i=(-1)^{\grade{i}\grade{g}}(f\rde_i)g+f(g\rde_i)~.$$
Morever we have
$$\grade{\lde_i\Phi}=\grade{\rde_i\Phi}\stackrel{\mathrm{mod}2}{=}
\grade{\Phi}+\grade{i}~,\qquad
\lde_j\lde_i=(-1)^{\grade{i}\cdot{\grade{j}}}\lde_i\lde_j~.$$

In an already quoted paper~\cite{C14b}
we considered a slightly different definition of partial derivatives,
tailored to the purpose of recovering, in the quantum setting,
usual coordinate expressions in Lagrangian field theory
formulated on jet bundles.
We observed that if \hbox{$v=v^i\,\de_i:\EC\to\VO\EC$}
is a vertical vector field,
then the Lie derivative $v.f$ is well-defined for any \hbox{$f:\M\to\FC$}
and obeys the standard Leibnitz rule.
Accordingly we used the setting \hbox{$v.f=(\de_if)\,v^i$},
which for \hbox{$\grade{v}=\grade{v^i}=\grade{i}$} yields
$$v.f=v^i\,\lde_i f=(-1)^{\grade{v}}(f\rde_i)\,v^i \qRq
\de_i=(-1)^{\grade{i}}\rde_i~.$$
\remark~Usually, in the literature,
one rather finds the convention of writing \hbox{$\de_i\equiv\lde_i$}\,.

\subsection{Batalin-Vilkovisky algebra}
\label{ss:Batalin-Vilkovisky algebra}

Recalling the quantum bundles introduced at the beginning
of~\sref{ss:Quantum bundles and quantum polynomials},
and assuming that quantum fields are sections \hbox{$\M\to\EC$},
we may enlarge our theory by considering \emph{antifields},
that is sections
\hbox{$\M\to\td\EC{}^*=\td\EC{}^*_\gradezero\dir{\M}\td\EC{}^*_\gradeone$} where
$$\td\EC{}^*_\gradezero\equiv\OC_\gradezero\tn\E^*_\gradeone~,\qquad
\td\EC{}^*_\gradeone\equiv\OC_\gradeone\tn\E^*_\gradezero~.$$
If we allow for all possible fields in these new sectors,
then for each field we have an antifield with inverted parity
and index position.
We note that the terminology can be a little confusing,
since the notion of antifield in this sense is not related
to that of anti-particle, and the anti-ghost is not the ghost's antifield.
Starting from the setting
of~\sref{ss:Pre-quantum fields of an essential gauge theory},
for example, in principle we might get an antifield for each one
of the fields $\gh$\,, $\bar\gh$ and $n$\,.
By the way, in that case the sectors of $\E$ include two copies of $\Lie$,
one fermionic and one bosonic, and one fermionic copy of $\Lie^*$;
so we note that things can get somewhat mixed-up.

The linear fiber coordinates $\bigl(y^i\bigr)$ determine
the dual fiber coordinates $\bigl(y_i\bigr)$ on $\E^*$.
When we see these as $\OC$-valued coordinates on $\td\EC{}^*$
we can, for clarity, denote them as $\bigl(\td y_i\bigr)$.
Accordingly, the antifield corresponding to the field $\phi$
will be denoted as \hbox{$\td\phi=\td\phi_i\,\de\td y^i$}.
The extended $\ZZ_2$-graded algebra of fiber polynomials
with numeric coefficients on $\EC\oplus\td\EC{}^*$ is
\begin{align*}
\FC'\equiv\FC\otimes\td\FC{}^*&\cong
{\vee}\E^*_\gradezero\otimes{\wedge}\E^*_\gradeone\otimes
{\wedge}\E_\gradezero\otimes{\vee}\E_\gradeone\cong
\\[6pt]
&\cong {\vee}\bigl(\E^*_\gradezero\oplus\E_\gradeone\bigr)\otimes
{\wedge}\bigl(\E^*_\gradeone\oplus\E_\gradezero\bigr)
\end{align*}
Here we have the partial derivatives $\lde_i$ and $\rde_i$
as in~\sref{ss:Quantum bundles and quantum polynomials},
and analogously defined derivatives $\ldea^i$ and $\rdea^i$
with similar properties.
But note that
$$1+\grade{i}\stackrel{\mathrm{mod}2}{=}\grade{\td i}
\equiv\grade{\td y^i}=\grade{\td y_i}~.$$

Next we consider the identity section
\hbox{$\id:\M\to\E\tn\E^*\cong(\E_\gradezero\tn\E^*_\gradezero)
\oplus(\E_\gradeone\tn\E^*_\gradeone)$}\,.
Its coordinate expression is \hbox{$\id=y_i{\sst\tn}y^i$}\,,
so that for \hbox{$f,g\in\FC'\equiv\FC\otimes\td\FC{}^*$} we get
$$\bang{\id\,|\,f}=\lde_i\ldea^i f~,\qquad
\bang{f\,|\,\id^*\,|\,g}=( f\rdea^i)\swe(\lde_i g)~,$$
where \hbox{$\id^*=y^i{\sst\tn}y_i$} is the transpose of $\id$\,.
Accordingly, we introduce the following maps.
\begin{definition}~\\
$\bullet$~The \emph{Batalin-Vilkovisky Laplacian} is the linear map
$$\Delta:\FC'\to\FC':f\mapsto\bang{\id\,|\,f}~.$$
\smallbreak\noindent
$\bullet$~The \emph{Batalin-Vilkovisky bracket} is the bilinear map
\hbox{$\{\_\,,\_\}:\FC'\times\FC'\to\FC'$}
which on $\ZZ_2$-homogeneous elements acts as
$$\{f,g\}\equiv\bang{f\,|\,\id^*\,|\,g}
-(-1)^{(\grade{f}+1)(\grade{g}+1)}\bang{g\,|\,\id^*\,|\,f}~.$$
\end{definition}

Then, by means of coordinate calculations, it's not difficult to prove:
\begin{proposition}
We have \hbox{$\Delta\!^2=0$}\,.
Moreover if \hbox{$f,g,h\in\FC'$} then we have
\begin{align*}
&\Delta(f\swe g)=\Delta f\swe g+(-1)^{\grade{f}}\{f,g\}
+(-1)^{\grade{f}}f\swe\Delta g~,
\\[6pt]
&\{f,g\swe h\}=\{f,g\}\swe h+(-1)^{(\grade{f}+1)(\grade{g}+1)}g\swe\{f,h\}~,
\end{align*}
provided that $f$ and $g$ are $\ZZ_2$-homogeneous.
\end{proposition}

The latter is a Jacobi-type identity,
and can be expressed as saying that the linear map
\hbox{$\ad_f\equiv\{f,\_\}:\FC'\to\FC'$}
turns out to be an anti-derivation of grade \hbox{$\grade{f}+1$}\,.

\subsection{BRST symmetry in Lagrangian field theory}
\label{ss:BRST symmetry in Lagrangian field theory}

It's not difficult to show that since \hbox{$\EC\onto\M$} is a vector bundle,
the $k$-jet bundle \hbox{$\JO_k\EC\onto\M$}
is also a vector bundle\footnote{
While \hbox{$\JO_k\EC\onto\JO_{k-1}\EC$} is always an affine bundle,
independently of any algebraic structure
in the fibers of \hbox{$\EC\onto\M$}.} 
\hbox{$\forall\,k\in\NN$}\,.
If $\bigl(x^\l,y^i\bigr)$ are fibered coordinates on $\E$,
we indicate the induced fiber coordinates on $\JO_k\E$
as $\bigl(y^i_{\sst\Lambda}\bigr)$,
where $\scriptstyle\Lambda$ is a multi-index of length
\hbox{$0\leq|{\scriptstyle\Lambda}|\leq k$}\,.
These can be seen as $\OC$-valued coordinates
on \hbox{$\JO_k\EC\cong\OC\tn\JO_k\E$}.
Now we observe that the notions and results
of~\sref{ss:Quantum bundles and quantum polynomials}
and of~\sref{ss:Batalin-Vilkovisky algebra}
can be straightforwardly extended by replacing $\EC$ with $\JO_k\EC$.
The related spaces of polynomials will be labeled by a subscript $k$\,,
so that we'll write \hbox{$\FC'_{\!k}\equiv\FC_{\!k}^{}\otimes\td\FC{}^*_{\!k}$}\,.

There exists a large
literature~\cite{MaMo83b,KrupkaSanders08,KrupkaKrupkovaSanders10}
about the jet bundle formulation
of Lagrangian field theories and their symmetries.
An approach consistent with the notion of quantum bundle presented here
was proposed in a previous paper~\cite{C14b};
we refer to it for further citations and mathematical details.
Two basic notions in that context are the notion of a ``totally horizontal form''
\hbox{$\a:\JO_k\EC\to\OC\tn\weu{q}\TS\M$}
and of an ``infinitesimal vertical transformation''
\hbox{$v:\JO\EC\to\VO\EC$},
required to be F-smooth morphisms over $\M$ and over $\EC$, respectively.
One introduces a natural operation
$$\d[v]\a\equiv\LO[v_{\sst(k)}]\a:\JO_{k+1}\EC\to\OC\tn\weu{q}\TS\M$$
defined as the Lie derivative of the basic form $\a$ along the
holonomic $k$-jet prolongation $v_{\sst(k)}$ of $v$
($\a$ can be viewed as a form on $\JO_k\EC$).
In particular we are interested in considering a first-order Lagrangian density
\hbox{$\Lcal=\ell\,\dO^4x:\JO\EC\to\OC\tn\weu4\TS\M$}.
Writing \hbox{$v=v^i\,\de_i$} and assuming
\hbox{$\ell,v^i:\JO\EC\to\OC$} to be fiber polynomials with numeric coefficients,
namely \hbox{$\ell,v^i:\M\to\FC_{\!1}^{}$}\,,
we obtain the coordinate expression
$$\d[v]\Lcal=
\bigl(v^i\,\lde_i\ell+\dO_\l v^i\,\lde^\l_i\ell\bigr)\,\dO^4x~,$$
where \hbox{$\dO_\l v^i\equiv\de_\l v^i+y^j\,\lde_j v^i+y^j_\l\,\lde_j^\l v^i$}
are the components of the horizontal differential\footnote{
A map \hbox{$f:\JO_k\EC\to\OC$} can be viewed as a basic $0$-form.
It's horizontal differential \hbox{$\dH f:\JO_{k+1}\EC\to\OC\tn\TS\M$}
is characterized by the property that for any F-smooth section
\hbox{$\phi:\M\to\EC$} one has
\hbox{$\dH f\comp\jO_{k+1}\phi=\dO(f\comp\jO_k\phi)$}\,.} 
$\dH v^i$.
One says that $v$ is an \emph{infinitesimal vertical symmetry}
if $\d[v]\Lcal$ is a horizontal differential, that is
\hbox{$\d[v]\Lcal=\dH\Ncal$} with
\hbox{$\Ncal:\JO\EC\to\OC\tn\weu3\TS\M$}.
A generalized version of the Noether theorem then follows,
as it turns out that \hbox{$v^i\,\lde_i^\l\ell\,\dO x_\l-\Ncal$}
is a \emph{conserved current}\,,
namely its evaluation through a "critical field"
(a solution of the Euler-Lagrange equations) yields a closed 3-form.

The usual BRST symmetry of a gauge field theory
can be expressed in terms of the operator $\d[v]$
with \hbox{$v=\th\,\breve v=\th\,\breve v^i\,\de y_i$}\,,
where \hbox{$\th\in\OC_\gradeone$} is any fixed odd element;
the components $\breve v^i$ are certain assigned fiber polynomials
\hbox{$\M\to\FC_{\!1}^{}$} which have \emph{inverted parity}
with respect to their respective sectors, that is
\hbox{$\grade{\breve v^i}\smash{\stackrel{\mathrm{mod}2}{=}}1+\grade{i}$}\,.
Then one defines the BRST transformation $\brstS$\,,
acting on totally horizontal forms, by \hbox{$\d[v]\a\equiv\th\,\brstS\a$}\,.
On fiber polynomials (seen as $0$-forms) $\brstS$ act as an anti-derivation.
In particular we have \hbox{$\breve v^i=\brstS y^i$}.

\begin{remark}
Many physics texts do not use the jet bundle approach
to Lagrangian field theories, but deal with a functional approach
involving the action integral.
Accordingly, the BRST transformation can be expressed as the ``Slavnov operator''
\hbox{$\breve v^{\sst X}\d/\d\phi^{\sst X}$}\,, where \hbox{${\scriptstyle X}\equiv(x,i)$}
is a generalized index including ordinary indices and spacetime position
and $\d/\d\phi^{\sst X}$ is the functional derivative
(\eg\ see Weinberg~\cite{We96},~Ch.15).
Summation with respect to position is to be intended as integration,
as in~\sref{ss:Generalized semi-densities and quantum states}.
\end{remark}\smallbreak

Since we deal with linear fiber coordinates
we can write \hbox{$\de y_i\equiv y_i$} and
\hbox{$\breve v=\breve v^i\,y_i$}\,,
which can also be regarded as the fiber polynomial
$$\breve v=(\brstS y^i){\sst\tn}\td y_i\in\FC_{\!1}^{}\tn\td\FC{}_{\!0}^*~.$$
Accordingly we may consider a Lagrangian density
of the form \hbox{$\Lcal=\Lcal_0+\breve v\,\eta$}\,,
where $\Lcal_0$ is the Lagrangian density of matter and gauge fields
and $\eta$ is the spacetime volume form.
This could be viewed as related to an extended theory
containing fields \emph{and} antifields,
but the essential idea is somewhat different:
we assign an ``antifield map'' map \hbox{$\s:\FC_{\!1}^{}\to\td\FC{}_{\!0}^*$}
such that \hbox{$\breve v\comp\s=(\brstS y^i){\sst\swe}\s_i$} is $\brstS$-exact
up to a horizontal differential
(where \hbox{$\s_i\equiv\td y_i\comp\s$}).
Since $\Lcal_0$ is $\brstS$-exact by construction, $\Lcal$ is $\brstS$-exact
up to a horizontal differential,
as we'll see in the example worked out in~\sref{ss:The fundamental example}.

\subsection{The fundamental example}
\label{ss:The fundamental example}

In~\sref{s:Quantum fields in a gauge theory}
we discussed the quantum free fields of an essential gauge theory.
We now look at some aspects of the corresponding theory of interacting fields,
seen as sections of a quantum bundle of the kind considered
in~\sref{ss:Quantum bundles and quantum polynomials}.
A jet bundle Lagrangian formulation of this theory
was previously examined~\cite{C14b}.
Here we'll recall some needed results and relate them to antifields
according to the scheme proposed
in~\sref{ss:BRST symmetry in Lagrangian field theory}.

Recalling~\sref{ss:Pre-quantum fields of an essential gauge theory}
we have the fermionic fields $\psi,\bar\psi,\gh,\bar\gh$,
and the bosonic fields $A$, $n$.
Now we face the following notational problem:
for clarity, we'd rather indicate fiber coordinates and field components
by different letters, as in \hbox{$\phi^i\equiv y^i\comp\phi$}
(\sref{ss:Quantum bundles and quantum polynomials});
but we now drop that distinction, like most physics texts,
as it is impractical when several sectors are involved;
namely we write $\phi^i$ for $y^i$.
Usually, the context should made things clear.
In particular, the components \hbox{$\s_i\equiv\td y_i\comp\s$}
of the ``antifield map'' map \hbox{$\s:\FC_{\!1}^{}\to\td\FC{}_{\!0}^*$}
(introduced in~\sref{ss:BRST symmetry in Lagrangian field theory})
will be expressed as $\td\phi_i$\,.
Actually by availing of a fiber metric
one can possibly associate, via $\s$\,,
an antifield \hbox{$\td\phi:\M\to\smash{\td\EC^*}$}
with each field \hbox{$\phi:\M\to\EC$}.

Henceforth we'll drop the symbol ${\scriptstyle\lozenge}$
and denote the $\ZZ_2$-graded product of fiber polynomials
by simple juxtaposition.

\smallbreak
We write the matter fermion~\&~gauge field Lagrangian as
\hbox{$\Lcal_0=(\ell_\psi+\ell_A)\,\dO^4x$} with
$$\ell_\psi=\bigl(
\ih\,(\bar\psi_{\a i}\,\nasl\psi^{\a i}
-\nasl\bar\psi_{\a i}\,\psi^{\a i})
-m\,\bar\psi_{\a i}\,\psi^{\a i}\bigr) \rdg~,\qquad
\ell_A=-\tfrac14\,g^{\l\m}\,g^{\n\r}\,F\iI{\l\n}\sI\,F_{\m\r\sI}\,\rdg~,$$
where $\nasl$ is the Dirac operator, $F$ is the curvature tensor of $A$
and $\rdg\dO^4x$ is the coordinate expression of the spacetime volume form $\eta$\,.
The BRST transformation is determined, according to the procedure
sketched in~\sref{ss:BRST symmetry in Lagrangian field theory}, by
\begin{align*}
&\breve v=
\breve v^{\a i}\,\dde{}{\psi^{\a i}}+\breve v_{\a i}\,\dde{}{\bps_{\a i}}
+\breve v^\sI_\l\,\dde{}{A_\l^\sI}
+\breve v^\sI\,\dde{}{\gh^\sI}+\breve v_\sI\,\dde{}{\bar\gh_\sI}~,
\\[6pt]
&\breve v^{\a i}=\lfr\iIi\sI ij\,\gh^\sI\,\psi^{\a j}\,,~
\breve v_{\a i}=\lfr\iIi \sI ji\,\bps_{\a j}\,\gh^\sI\,,~
\breve v^\sI_\l=\gh^\sI_{;\l}\,,~
\breve v^\sI=\oh\,\sco\sI\sJ\sH\,\gh^\sJ\,\gh^\sH\,,~
\breve v_\sI=n_\sI\,,
\end{align*}
with
\hbox{$\gh^\sI_{;\l}\equiv\na_\l\gh\equiv\gh^\sI_{,\l}
+\oh\,\sco\sI\sJ\sH\,\gh^\sJ\,A^\sH_\l$}\,.
Then it's not difficult to check that \hbox{$\brstS^2=0$} as expected.

\begin{remark}
The first three terms in $\breve v$ determine the action of $\brstS$
on the matter~\&~gauge Lagrangian $\Lcal_0$\,,
which is exactly an infinitesimal gauge transformation parametrized by $\gh$\,.
Hence (as it can also be checked by direct calculations)
we have \hbox{$\brstS\Lcal_0=0$}\,.
The geometrical nature of the fourth term is also interesting:
it is essentially the map \hbox{$\Lie^*\to\Lie^*{\sst\tn}\Lie^*$}
dual of the Lie algebra product \hbox{$[\_\,,\_]:\Lie{\sst\tn}\Lie\to\Lie$}\,.
The last term is given in that form in order to get \hbox{$\brstS^2=0$}
in all cases, but could be otherwise replaced by \hbox{$n_\sI=-f_\sI/\xi$}
that is the ``field equation'' derived from the total Lagrangian (below).
\end{remark}\smallbreak

In this context we consider a map $\s$ which has non-zero components
only in the antifield sectors corresponding to the gauge field $A$
and to the anti-ghost field $\bar\gh$\,, and is given by
$$\td A^\l_\sI=-g^{\l\m}\,\bar\gh_{\sI,\m}~,\qquad
\Tilde{\Bar\gh}{}^\sI=(f^\sI+\oh\,\xi\,n^\sI)~,$$
where \hbox{$\xi\in\RR$} and $f^\sI$ is a shorthand for
$\frac1\rrdg\,\dO_\l(g^{\l\m}\,\rdg\,A^\sI_\m)$\,.
Then, setting
$$\Kcal\equiv\bar\gh_\sI\,(f^\sI+\oh\,\xi\,n^\sI)\,\rdg\dO^4x~,\qquad
\Mcal\equiv g^{\l\m}\,\rdg\,\bar\gh_\sI\,\gh^\sI_{;\m}\,\dO x_\l~,$$
where \hbox{$\dO x_\l\equiv\de x_\l\pint\dO^4x$}\,,
by straightforward calculations we find
\begin{align*}
\Lcal\ghost\equiv(\breve v\comp\s)\,\eta&=
\bigl((\brstS A_\l^\sI)\,\td A^\l_\sI
+(\brstS\bar\gh_\sI)\,\Tilde{\Bar\gh}{}^\sI\bigr)\,\rdg\dO^4x=
\\[6pt]
&=\bigl(g^{\l\m}\,\bar\gh_{\sI,\l}\,\gh^\sI_{;\m}
+n_\sI\,(f^\sI+\oh\,\xi\,n^\sI)\bigr)\,\rdg=
\\[6pt]
&=\brstS\Kcal+\dH\Mcal~.
\end{align*}

\begin{remark}
The assignment of $\Tilde{\Bar\gh}{}^\sI$,
and in particular of the constant $\xi$,
can be viewed as the fixing of a gauge condition.
\end{remark}\smallbreak

We now consider the extended Lagrangian \hbox{$\Lcal=\Lcal_0+\Lcal\ghost$}\,,
which is still first-order since $\Lcal\ghost$ is such.
We note, however, that the equivalent Lagrangian \hbox{$\Lcal'=\Lcal_0+\brstS\Kcal$}
is second-order, since the two terms $\brstS\Kcal$ and $\dH\Mcal$ are such.
On the other hand, $\Lcal$ and $\Lcal'$ are both BRST-invariant,
in a generalized sense, and give rise to the same BRST current.
In order to prove these claims we use
a formulation of the Noether theorem applicable in the present context~\cite{C14b},
of which the first-order situation briefly summarized
in~\sref{ss:BRST symmetry in Lagrangian field theory}
is a special case.

Let \hbox{$\Lcal=\ell\,\dO^4x:\JO_k\EC\to\OC\tn\weu4\TS\M$}
be a $k$-order Lagrangian; a morphism 
\hbox{$v:\JO\EC\to\VO\EC$} over $\EC$
is called an \emph{infinitesimal vertical symmetry} of $\Lcal$
if there exists \hbox{$\Ncal:\JO_k\EC\to\OC\tn\weu3\TS\M$}
such that \hbox{$\d[v]\Lcal=\dH\Ncal$}.
In such case we obtain a \emph{conserved current}
$$\Jcal=\Jcal^\l\,\dO x_\l:\JO_{2k-1}\EC\to\OC\tn\weu3\TS\M$$
with the expression
$$\Jcal^\l=v_{\sst(k-1)}|\Pcal^\l-\Ncal^\l~,$$
where $\Pcal^\l$ is a certain morphism of order \hbox{$2k-1$} which,
in general, has a rather complicate expression.
For \hbox{$k=2$} we obtain the relatively simple expression
$$v_{\sst(1)}|\Pcal^\l=v^i\,(\lde^\l_i\ell-\dO_\m\lde^{\l\m}_i\ell)
+\dO_\m v^i\,\lde^{\l\m}_i\ell~.$$

Coming back to the ghost Lagrangian and the BRST symmetry
the argument goes now as follows.
First we note that \hbox{$\d[v]\Lcal\ghost=\dH\d[v]\Mcal$}\,,
since \hbox{$\d[v]\brstS\Kcal=\th\,\brstS^2\Kcal=0$}\,,
so that $\d[v]\Mcal$ plays the role of $\Ncal$.
Then a straightforward calculation shows that indeed $\d[v]\Mcal$
equals the part of $v_{\sst(1)}|\Pcal^\l$ which derives from $\dH\Mcal$,
so that we may actually conclude that adding a term proportional to $\dH\Mcal$
does not alter the BRST-invariance of a Lagrangian nor the related current.

Our final remark concerns the relation between BRST charge
and equal-time super-commutators.
If $v$ is an arbitrary infinitesimal vertical symmetry
and $\Jcal$ is the corresponding current then~\cite{C14b}
$$\brstQ\equiv\int\dO^3x_\sbo\,\Jcal^0(x)=
\int\dO^3x_\sbo\,\p_i(x)\,v^i\comp\jO\phi(x)$$
is constant when evaluated through critical sections.
If moreover $\Jcal^0$ has even parity
and \hbox{$\Suc{\phi^i(x)\,,\,v^j[\phi](x')}=0$} at equal times,
then it's not difficult to see that
$$\d[v]\phi^i=\iO\,\bigl[\brstQ\,,\phi^i\bigr]\,\detg^{-1/2}~.$$
In the case of the BRST symmetry the charge in the above sense
is actually $\th\brstQ$\,,
and eliminating $\th$ we obtain
$$\brstS\phi^\a=\iO\,\Suc{\brstQ\,,\,\phi^\a}\,\detg^{-1/2}~.$$
The validity of this relies on the condition
that the super-commutators
\hbox{$\Suc{\phi^i(x)\,,\,v^j[\phi](x')}$} vanish at equal times in all sectors.
This condition is indeed fulfilled for all free fields
explicitely constructed as described
in~\sref{s:Quantum fields in a gauge theory};
its validity for fully interacting critical fields
is usually inferred by general arguments based on the form of the dynamics.


\end{document}